\PassOptionsToPackage{unicode}{hyperref}
\PassOptionsToPackage{hyphens}{url}
\PassOptionsToPackage{dvipsnames,svgnames,x11names}{xcolor}
\documentclass[
  12pt]{article}

\usepackage{shortex} 
\RequirePackage{amsthm,amsmath,amsfonts,amssymb}
\RequirePackage[numbers]{natbib}
\usepackage{forest}
\usepackage{tikz}

\usetikzlibrary{shapes,decorations,arrows,calc,arrows.meta,fit,positioning}
\tikzset{
    -Latex,auto,node distance =1 cm and 1 cm,semithick,
    state/.style ={circle, draw, minimum width = 0.75cm},
    point/.style = {circle, draw, inner sep=0.04cm,fill,node contents={}},
    bidirected/.style={Latex-Latex,dashed},
    el/.style = {inner sep=2pt, align=left, sloped}
}

\usepackage{amsmath,amssymb}
\usepackage{iftex}
\ifPDFTeX
  \usepackage[T1]{fontenc}
  \usepackage[utf8]{inputenc}
  \usepackage{textcomp} 
\else 
  \usepackage{unicode-math}
  \defaultfontfeatures{Scale=MatchLowercase}
  \defaultfontfeatures[\rmfamily]{Ligatures=TeX,Scale=1}
\fi
\usepackage{lmodern}
\ifPDFTeX\else  
\fi
\IfFileExists{upquote.sty}{\usepackage{upquote}}{}
\IfFileExists{microtype.sty}{
  \usepackage[]{microtype}
  \UseMicrotypeSet[protrusion]{basicmath} 
}{}
\makeatletter
\@ifundefined{KOMAClassName}{
  \IfFileExists{parskip.sty}{%
    \usepackage{parskip}
  }{
    \setlength{\parindent}{0pt}
    \setlength{\parskip}{6pt plus 2pt minus 1pt}}
}{
  \KOMAoptions{parskip=half}}
\makeatother
\usepackage{xcolor}
\makeatletter
\ifx\paragraph\undefined\else
  \let\oldparagraph\paragraph
  \renewcommand{\paragraph}{
    \@ifstar
      \xxxParagraphStar
      \xxxParagraphNoStar
  }
  \newcommand{\xxxParagraphStar}[1]{\oldparagraph*{#1}\mbox{}}
  \newcommand{\xxxParagraphNoStar}[1]{\oldparagraph{#1}\mbox{}}
\fi
\ifx\subparagraph\undefined\else
  \let\oldsubparagraph\subparagraph
  \renewcommand{\subparagraph}{
    \@ifstar
      \xxxSubParagraphStar
      \xxxSubParagraphNoStar
  }
  \newcommand{\xxxSubParagraphStar}[1]{\oldsubparagraph*{#1}\mbox{}}
  \newcommand{\xxxSubParagraphNoStar}[1]{\oldsubparagraph{#1}\mbox{}}
\fi
\makeatother

\usepackage{longtable,booktabs,array}
\usepackage{calc} 
\usepackage{etoolbox}
\makeatletter
\patchcmd\longtable{\par}{\if@noskipsec\mbox{}\fi\par}{}{}
\makeatother
\IfFileExists{footnotehyper.sty}{\usepackage{footnotehyper}}{\usepackage{footnote}}
\makesavenoteenv{longtable}
\usepackage{graphicx}
\makeatletter
\def\maxwidth{\ifdim\Gin@nat@width>\linewidth\linewidth\else\Gin@nat@width\fi}
\def\maxheight{\ifdim\Gin@nat@height>\textheight\textheight\else\Gin@nat@height\fi}
\makeatother
\setkeys{Gin}{width=\maxwidth,height=\maxheight,keepaspectratio}
\makeatletter
\def\fps@figure{htbp}
\makeatother

\makeatletter
\@ifpackageloaded{caption}{}{\usepackage{caption}}
\AtBeginDocument{%
\ifdefined\contentsname
  \renewcommand*\contentsname{Table of contents}
\else
  \newcommand\contentsname{Table of contents}
\fi
\ifdefined\listfigurename
  \renewcommand*\listfigurename{List of Figures}
\else
  \newcommand\listfigurename{List of Figures}
\fi
\ifdefined\listtablename
  \renewcommand*\listtablename{List of Tables}
\else
  \newcommand\listtablename{List of Tables}
\fi
\ifdefined\figurename
  \renewcommand*\figurename{Figure}
\else
  \newcommand\figurename{Figure}
\fi
\ifdefined\tablename
  \renewcommand*\tablename{Table}
\else
  \newcommand\tablename{Table}
\fi
}
\@ifpackageloaded{float}{}{\usepackage{float}}
\floatstyle{ruled}
\@ifundefined{c@chapter}{\newfloat{codelisting}{h}{lop}}{\newfloat{codelisting}{h}{lop}[chapter]}
\floatname{codelisting}{Listing}

\makeatother
\makeatletter
\@ifpackageloaded{caption}{}{\usepackage{caption}}
\@ifpackageloaded{subcaption}{}{\usepackage{subcaption}}
\makeatother

\ifLuaTeX
  \usepackage{selnolig}  
\fi
\usepackage[]{natbib}
\usepackage{bookmark}

\IfFileExists{xurl.sty}{\usepackage{xurl}}{} 
\hypersetup{
  pdftitle={Title},
  pdfauthor={Author 1; Author 2},
  pdfkeywords={3 to 6 keywords, that do not appear in the title},
  colorlinks=true,
  linkcolor={blue},
  filecolor={Maroon},
  citecolor={Blue},
  urlcolor={Blue},
  pdfcreator={LaTeX via pandoc}}

\newcommand{\anon}{1}

\definecolor{julia1}{rgb}{0, 0.60, 0.98} 
\definecolor{julia2}{rgb}{0.89, 0.43, 0.28}
\definecolor{julia3}{rgb}{0.24, 0.64, 0.30}
\definecolor{julia4}{rgb}{0.76, 0.44, 0.82}
\definecolor{julia5}{rgb}{0.67, 0.55, 0.09}
\definecolor{julia6}{HTML}{00aaae}
\definecolor{julia7}{HTML}{ed5e93}
\definecolor{julia8}{HTML}{c68225}

\algrenewcommand\algorithmicrequire{\textbf{Input:}}
\algrenewcommand\algorithmicensure{\textbf{Output:}}

\DeclareMathOperator{\logistic}{logistic}

\newcommand{\forwardAD}{\texttt{forwardAD}}
\newcommand{\reverseAD}{\texttt{reverseAD}}
\newcommand{\JacobianAD}{\texttt{JacobianAD}}
\newcommand{\HessianAD}{\texttt{HessianAD}}
\newcommand{\CGM}{\texttt{CG\_M}}
\newcommand{\CGH}{\texttt{CG\_H}}
\newcommand{\BlindM}{\texttt{Unaware\_M}}
\newcommand{\BlindH}{\texttt{Unaware\_H}}
\newcommand{\M}{\texttt{M}}
\newcommand{\Hfunc}{\texttt{H}}
\newcommand{\CGGradH}{\texttt{CG\_Grad\_H}}
\newcommand{\BlindGradH}{\texttt{Unaware\_Grad\_H}}
\newcommand{\NaiveGradH}{\texttt{Naive\_Grad\_H}}

\newcommand{\leapfrog}{\Phi_\text{leapfrog}}
\newcommand{\kick}{\Phi_\text{kick}}
\newcommand{\drift}{\Phi_\text{drift}}
\newcommand{\midpoint}{\Phi_\text{midpoint}}
\newcommand{\hmc}{\Phi_\text{HMC}}
\newcommand{\rhmc}{\Phi_\text{RHMC}}

\newcounter{xxx}
\renewcommand{\paragraph}[1]{\noindent{\bf #1}}

\begin{document}

\def\spacingset#1{\renewcommand{\baselinestretch}%
{#1}\small\normalsize} \spacingset{1}


\if1\anon
{
  \title{\bf Hessian-informed, Coordinate Friendly Hamiltonian Monte Carlo in Linear Time}
  \author{Son Luu \and Nikola Surjanovic \and Zuheng Xu \and
  Trevor Campbell \and Alexandre Bouchard-C\^{o}t\'{e}
  \thanks{
    ABC and TC acknowledge the support of a CANSSI CRT Grant and NSERC Discovery Grants. 
    NS and ZX notes that this work does not relate to their current positions at Amazon.
    We additionally acknowledge use of the ARC Sockeye computing platform from the University of British Columbia.}\hspace{.2cm}\\
    Department of Statistics, The University of British Columbia}
  \maketitle
} \fi

\if0\anon
{
  \bigskip
  \bigskip
  \bigskip
  \begin{center}
    {\LARGE\bf Hessian-informed, Coordinate Friendly Hamiltonian Monte Carlo in Linear Time}
\end{center}
  \medskip
} \fi

\bigskip
\begin{abstract}
Riemannian Hamiltonian Monte Carlo (RHMC) is a promising MCMC methodology thanks to its ability to accommodate position-dependent preconditioning and multi-step proposals.
While RHMC performs well in low dimensions, it becomes infeasible in high dimensions due to its $O(d^3)$ cost per fixed-point iteration, where $d$ is 
the dimension of the target density. 
Even when the position-dependent preconditioner is based on the diagonal of the Hessian, the cost is still $O(d^2)$ per fixed-point iteration. 
In this paper, we propose a computational 
method to reduce the computational complexity of RHMC fixed-point iterations with diagonal preconditioners from $O(d^2)$
to $O(d)$ for targets with ``coordinate friendly'' structures. This distribution class 
includes generalized linear models as well as other dense and sparse graphical models. The method is expressed as manipulating the compute graph and can therefore be automated to work on black box targets.   Finally, we show empirically that our implementation of RHMC results in better sample quality per unit of compute time  for various target distributions 
compared to state-of-the-art HMC NUTS algorithms with both position-independent and 
position-dependent preconditioners.
\end{abstract}

\noindent%
\vfill

\newpage
\spacingset{1.8} 

\section{Introduction}

Hamiltonian Monte Carlo (HMC), introduced by \citet{duane1987hybrid}, is a class of Markov Chain Monte Carlo (MCMC)
methods making use of the Hamiltonian dynamics to obtain approximate samples of an unnormalized 
target distribution $\pi$. By augmenting
the state space with a momentum variable and performing several steps per proposal, HMC can perform bigger jumps compared to single-step methods such as  the 
Metropolis-adjusted Langevin  (MALA) or random-walk Metropolis–Hastings (MH) algorithms. 
For example, on ill-conditioned normal targets, the compute time needed to attain a fixed sample quality for randomized 
HMC can scale as the square root of the condition number \citep{langmore_condition_2019} (often formalized as the maximum square ratio of the broadest direction scale to that of the most constrained) while the compute time to a fixed accuracy scales at least linearly for most other existing samplers (for a recent review, see \citep[Section~4]{luu2025gibbs}).

\begin{figure}[t]
	\centering
	\begin{subfigure}{0.325\textwidth}
		\centering
		\includegraphics[width=\textwidth]{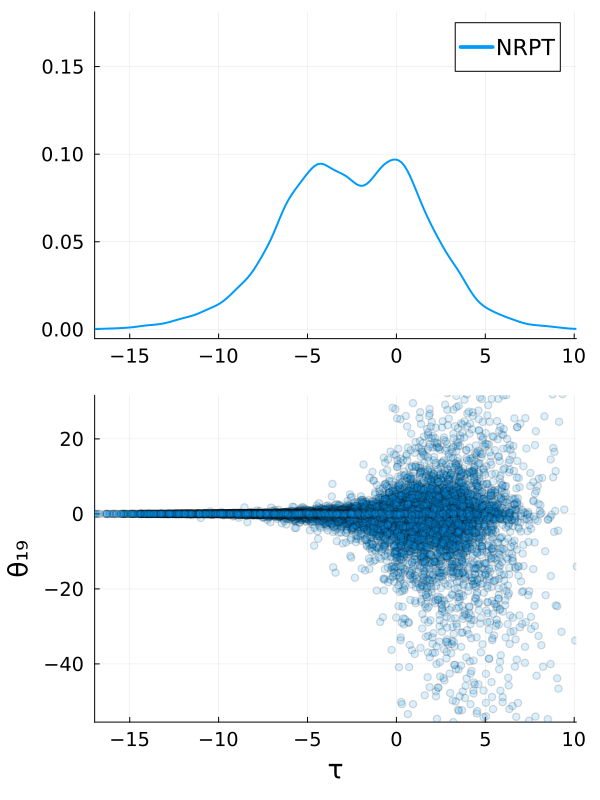}
	\end{subfigure}
	\begin{subfigure}{0.325\textwidth}
		\centering
		\includegraphics[width=\textwidth]{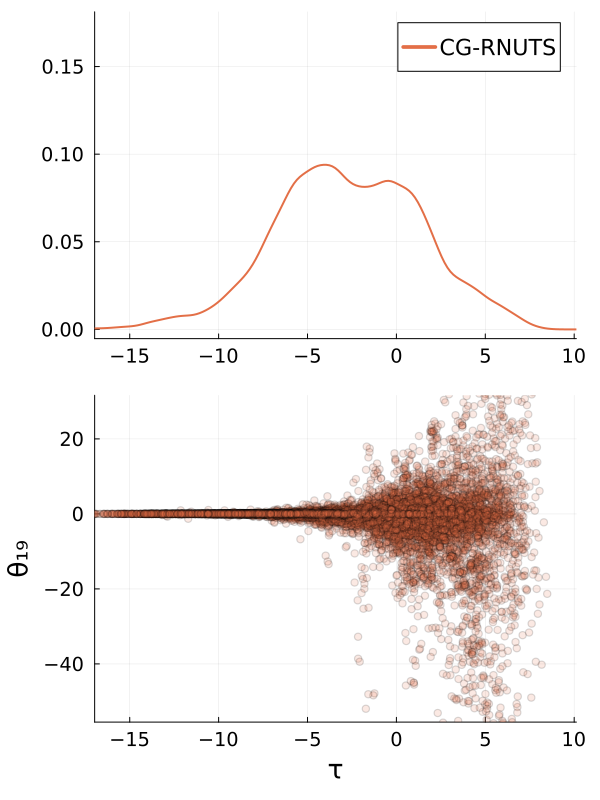}
	\end{subfigure}
	\begin{subfigure}{0.325\textwidth}
		\centering
		\includegraphics[width=\textwidth]{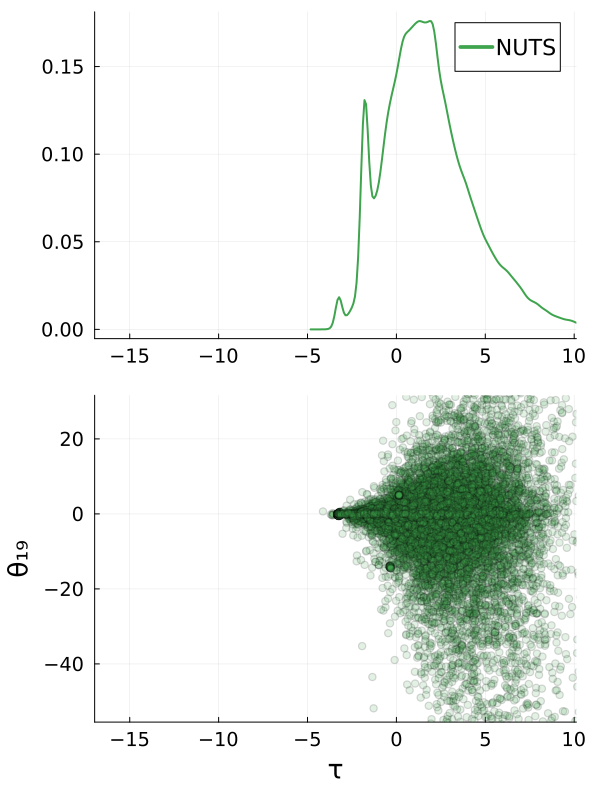}
	\end{subfigure}
	\caption{
		Visualization of samples obtained by NUTS and our CG-RNUTS method with similar wall clock time budgets. We also show for  comparison  a reference posterior obtained by non-reversible parallel tempering with a slice sampling explorer (NRPT). 
		All samplers target the same posterior distribution, coming from a horseshoe logistic regression model applied to prostate cancer data (Section~\ref{sec:realdata_experiment}). 
		The plots demonstrate the presence of a funnel-shaped posterior that is challenging to explore for standard  HMC NUTS methods (by ``standard'', we mean that the preconditioning mass matrix is position-independent).
		 \textbf{Bottom row}: bivariate marginal of the global scale parameter versus the 19$^{th}$ coefficient parameter (randomly chosen). 
		 \textbf{Top row}: marginal density of the global scale parameter. }
	\label{fig:horseshoe_marginals}
\end{figure}

HMC still struggles when the target has varying geometries
such as the \emph{funnel structure} \citep[Section 8]{neal_slice_2003} found in many hierarchical problems (see Figure~\ref{fig:horseshoe_marginals} for an example arising from real data), where the condition number is extremely large or even infinite. The root of the problem is that HMC relies on
a \emph{fixed} (position-independent) metric in the form of a positive-definite mass matrix to govern its momentum distribution, and this inflexible, fixed metric makes exploring high curvature
regions difficult and low curvature regions inefficient. To address this issue, generalizations of HMC where the metric can depend on the position have been developed in the physics \citep{creutzglobal1988} and statistics literature  \citep{girolami2011riemann}. These methods use position-dependent metrics to efficiently traverse regions of high and low curvatures. Similarly to second-order methods in optimization, this allows samplers to not only take bigger steps in flat directions and smaller steps in steep directions, but also to make this assessment dynamically, based on the current point. In some contexts, this flexibility can lead to convergence rates that are independent of the condition number \cite{kook_condition-number-independent_2023}. 

The method presented in \citet{girolami2011riemann}  is known as Riemannian HMC due to a specific choice of metric, namely the Fisher information, but the method has been generalized to other choices of metrics (see Table~1 in \citet{kailas2026hierarchical} for a review). We will refer to position-dependent momentum HMC algorithms collectively as RHMC, as often done in the computational statistics literature. In this work, we focus on variants of RHMC using the diagonal Hessian or transformation of it, since they can be applied more broadly than methods based on the Fisher information metric. Specifically, Hessian-based metrics can be automated using Automatic Differentiation (AD) while the Fisher information involves an expectation and therefore model-specific mathematical derivations to compute exactly. Alternatively, one can approximate
the Fisher information matrix via Monte Carlo estimates, but this approach is often more expensive than computing the log density Hessian.

Practical implementations of RHMC are challenging for several reasons. First, each \emph{integrator step} (i.e., each individual step comprising a proposed multi-step trajectory) is typically designed so that the overall proposal is an 
involution that exactly preserves volume, and approximately preserves the Hamiltonian to achieve high Metropolis-Hastings acceptance rates without needing to compute the determinant of the involution. In the case of standard HMC, this can be done using the leapfrog integrator, which is an explicit update, i.e., an update that has a closed-form expression. 
For RHMC, because the momentum distribution is changing with the position, it is more challenging to construct such an update. 
Early work on RHMC  \citep{girolami2011riemann}  used the generalized leapfrog \cite[p. 81--87]{Leimkuhler_Reich_2005}, a fixed-point iterative process that 
has to be done \emph{within} each step of a multi-step proposal. 
Subsequent work has shifted focus to implicit midpoint methods, which are still iterative algorithms, 
but have been shown to substantially outperform the generalized leapfrog \citep{brofos2021evaluating,brofos2021numerical}. 

In this work, we focus on a second practical challenge, which is that a single iteration of these implicit methods 
(be it the midpoint, or the generalized leapfrog, or any other alternative \citep{tao2016explicit}) requires a Cholesky decomposition in the general case, and hence a computation cost per implicit solver iteration that scales as $O(d^3)$, where $d$ is the number of parameters. 
To reduce this cost, several authors \citep{betancourt2013general,xu2024practical,kailas2026hierarchical} have proposed using a diagonal metric, in particular, metrics based on the diagonal of the Hessian. However, even in the diagonal Hessian case, the cost per implicit solver iteration is high: there is evidence that computing the diagonal entries of a Hessian matrix has time complexity $\Omega(d^2)$ \citep[Section 6]{Martens2012Estimating}. 
Moreover, RHMC requires not only forming the Hamiltonian, but also taking its gradient. We explain in Section~\ref{sec:hess_trick} that if it is not done carefully, the cost of this gradient can be cubic in $d$, even in the diagonal case. In fact, several existing code bases 
have a cubic empirical scaling (see Section~\ref{sec:faster}).

To address this challenge, we focus on a class of posterior distributions called \emph{coordinate friendly} targets (CF). 
We provide several examples of popular statistical models in that class, including sparse graphical models. Coordinate friendliness goes beyond sparse graphical models, and includes in particular high-dimensional generalized linear models (GLMs). 
For CF targets, we show that the diagonal Hessian matrix can be computed in $O(d)$. 
Moreover, we show how the gradient of the Hamiltonian can also be computed in $O(d)$ for CF targets, using an innovative ``reverse-over-forward'' 
automatic differentiation strategy. In contrast, most existing mixed automatic differentiation strategies perform ``forward-over-reverse.'' 

We have implemented our method in the Julia programming language
as an extension to the popular \texttt{AdvancedHMC.jl} 
software package \citep{xu2020advancedhmc}. 
We provide numerical results showing that our implementation 
is correct, i.e., that it produces the same RHMC 
trajectories as existing implementations up to numerical precision 
(Figure~\ref{fig:trajectory}). 
The key advantage of our method is that in medium to high dimension problems, 
the computational cost to produce a proposal is much lower than 
existing implementations. For example, we measured a speed-up of more than 100,000-fold in a 2,048 dimensional target (Figure~\ref{fig:timing}). 
Thanks to this speed-up, we observe empirically that our method 
outperforms not only previous implementations of RHMC but also state-of-the-art, adaptively tuned fixed metric baselines. 
These findings also hold when both our method and the  baselines are combined with a No-U-Turn (NUTS) 
integrator stopping criterion and multinomial trajectory 
resampling (denoted respectively RNUTS and NUTS) \citep{hoffman_no-u-turn_2014,betancourt2017conceptual}.

\paragraph{Related work.}  Since its inception, there have been several attempts to reduce the cost of 
RHMC integration steps for general high-dimensional target distributions. 
For targets with sparse Hessian matrices, \citet{kleppe2018modified} devised a modified 
Cholesky decomposition to exploit sparsity and speed up computation. In contrast to our method, this approach relies on manual computation of the metric, and moreover its computational gains vanish for targets with dense Hessians. 
In a more general setting, \citet{hartmann2022lagrangian} constructed their metric 
by adding a rank-one matrix to the identity. This metric uses only gradient information, 
so while the cost per integration step is competitive with non-Riemannian methods, their experiments show that this method still lags behind non-Riemannian 
counterparts in terms of time-normalized mixing rate. Most recently, \citet{kailas2026hierarchical} modeled 
the RHMC metric after the hierarchical structure of the target, enabling their method 
to use an explicit integrator. This method, however, still requires additional adaptation of 
the metric—unlike typical tuning-free RHMC metrics—to maintain linear cost per integration step. 
There are also several lines of work tailored to specific target distributions, e.g., for certain Gaussian process models \citep{paquet2018efficient}, for self-concordant barriers arising from linearly constrained sampling \citep{kook2022sampling,gatmiry2024sampling}, or for spectral density estimation in the space of positive definite matrices \citep{holbrook2018geodesic}. 
In the same vein as the original, Fisher-based RHMC 
approach \citep{girolami2011riemann}, these methods can only be applied on a case-by-case basis, making them harder to automate via probabilistic programming languages. 
To summarize, existing linear-scaling RHMC methods require either analytical Fisher information, 
sparsity, or additional metric adaptation. Our method, in contrast, does not require 
any of these conditions.

Beyond RHMC, there are many other MCMC methods with position-dependent preconditioners such as Riemannian MALA \citep{girolami2011riemann}, position-dependent Random Walk Metropolis \cite{livingstone2021geometric}, and unadjusted schemes \cite{dalalyan_theoretical_2017}. However, most alternatives are single-step proposal methods unlikely to enjoy RHMC's favourable scaling in the residual condition number \citep{kook_condition-number-independent_2023}. One exception is Piecewise Deterministic MCMC methods (PD-MCMC) \cite{vanetti_piecewise-deterministic_2018,bouchard-cote_bouncy_2018,bierkens_zig-zag_2019}, as their discretizations can be viewed as multi-step methods. One PD-MCMC method with a position-dependent mass matrix has been developed in \citet{kleppe_log-density_2024}, where forward differentiation and sparsity are used for efficient computation of a metric. However, since the method is based on numerical generalized randomized Hamiltonian Monte Carlo (NGRHMC) \citep{kleppe_connecting_2022} rather than RHMC, the method does not track metric derivatives, and hence  reverse-over-forward AD is not used (see \citet[Eq. (15)]{kleppe_connecting_2022}). While NGRHMC is simpler to implement compared to RHMC, NGRHMC is not exactly $\pi$-invariant. A special case of position-dependent preconditioning is automatic step size tuning methods \citep{biron2024automala,chevallier2025towards,bou2025within}, which can be viewed as only letting the magnitude of the metric vary but not its directions. While these methods can handle changing geometries, they remain highly sensitive to the target distribution's condition number. 

The coordinate friendly property has long been studied \citep{peng2016coordinate} and is widely used 
in optimization algorithms such as coordinate descent \citep{nesterov2012efficiency,wright2015coordinate}. 
However, compared to  optimization, 
RHMC brings additional challenges, in particular the 
requirement to differentiate the metric, leading to the need 
to efficiently compute third-order derivatives, which 
we address in this work.
Moreover, while stochastic optimization can accommodate  unbiased 
estimates of the Hessian diagonal \citep{agarwal2017second,yao2021adahessian}, 
the effect of the error on these estimates has not been studied in
the RHMC literature, so we focus here on exact diagonal 
Hessian matrices in this work. 

\section{Background}

For the rest of the paper, let $\pi$ be the target distribution of interest 
on $\reals^d$ with density $\pi(\theta)\propto e^{-U(\theta)}$, for 
$\theta:=(\theta_1,\ldots,\theta_d)^\top\in\reals^d$, with respect 
to Lebesgue measure on $\reals^d$.
We assume that for all $\theta\in\reals^d$, $\nabla U(\theta)$ and
$\nabla^2 U(\theta)$ are well-defined. We overload the $\diag(\cdot)$
notation with two operations: one taking in a vector
and returning a diagonal matrix with the vector as its diagonal;
one taking in a square matrix and returning a diagonal matrix 
with the same diagonal as the input matrix. In other words,
\[
\diag(A) = \diag(A_{1,1},\ldots,A_{d,d}) := \bpmat 
A_{1,1} & 0 & \ldots & 0\\ 
0 & A_{2,2} & \ldots & 0\\
\vdots & \vdots & \ddots & \vdots \\ 
0 & 0 & \ldots & A_{d,d} \epmat
\]
where $A$ is a $d$ by $d$ matrix with entries $A_{i,j}$ for $i, j \in \{1, 2, \dots, d\}$. 

\subsection{Hamiltonian Monte Carlo}

The Hamiltonian Monte Carlo algorithm (HMC) is an MCMC algorithm based on the deterministic alternation of two kernels: a Metropolis-Hastings (MH) sampler with a proposal given by a certain approximation of Hamiltonian dynamics, and a Gibbs sampling algorithm on a \emph{momentum auxiliary variable} $p \in \reals^d$. We review both below (for pedagogical exposition, see \citet{neal2011hmc}).

\paragraph{First kernel: Metropolized Hamiltonian dynamics } The inspiration of this kernel comes from Hamiltonian dynamics, an ordinary differential equation (ODE) given by:
\[\label{eq:hamiltonian_dynamics}
\frac{d}{dt}\theta(t) =\nabla_p H(\theta(t),p(t)), \;\;\;
\frac{d}{dt}p(t) = -\nabla_\theta H(\theta(t),p(t)).
\]
Here, $H$ is a function called the Hamiltonian and is defined as
\[
H(\theta,p) = U(\theta) + \frac{1}{2}\log(|M|) + \frac{1}{2}p^\top M^{-1}p
\]
where in this section the metric $M$ is a \textit{fixed} $d\times d$ user-defined positive-definite matrix (we will  make it dynamic in the next section), $|M|$ is the 
determinant of $M$ and $\nabla_\theta H, \nabla_p H$ are the gradient vectors
with respect to the first and second argument of $H$, respectively.
The HMC algorithm and its RHMC generalizations are defined on the state space $z = (\theta, p) \in \reals^{2d}$. 
The invariant distribution of HMC is a product of the target and a normal distribution with covariance matrix $M$:
\[
\bar{\pi}(\theta,p) := \pi(\theta)\Norm(p|0,M) = e^{-H(\theta,p)}.
\]
At each HMC iteration, we build a proposal as follows. First, define the leapfrog operator, 
\[
\leapfrog &= \kick \circ \drift \circ \kick \\
\kick(\theta, p) &= \left(\theta, p - \frac{\eps}{2}\nabla U(\theta)\right) \\
\drift(\theta, p) &= \left(\theta + \eps M^{-1} p, p\right),
\]
which is an approximation to the Hamiltonian dynamics, \cref{eq:hamiltonian_dynamics}.
Second, define a deterministic proposal by iterating the leapfrog $T$ times followed by a flip, $F(\theta, p) = (\theta, -p)$:
\[\label{eq:hmc-involution}
\hmc = F \circ \leapfrog^T. 
\] 
Let $z^* = \hmc(z)$ denote the deterministic proposal output. Since $\hmc$ is an involution, i.e., $\hmc = \hmc^{-1}$, by \citet{tierney_note_1998}, the MH algorithm with the acceptance probability given below is invariant with respect to $\bar \pi$
\[\label{eq:accept} \min\left\{ 1,  \frac{\bar \pi(z^*)}{\bar \pi(z)} |\nabla \hmc|\right\} = \min\left\{ 1,  \frac{\bar \pi(z^*)}{\bar \pi(z)}\right\}, \]
where in the above we can drop the determinant thanks to  the fact that $\hmc$ is a composition of \emph{shear transformations} (i.e., where only one of $\theta$ or $p$ changes) and hence has a unit Jacobian.

\paragraph{Second kernel: momentum refreshment } In this second kernel, the momentum $p$ is resampled according to a $d$-dimensional multivariate normal distribution with mean zero and covariance matrix $M$.  

The alternation of the two kernels described above corresponds to the original HMC 
algorithm \citep{duane1987hybrid} where the trajectory length is fixed and only the
final state is proposed. Subsequent work identified the importance of randomizing the trajectory length \citep{mackenze_improved_1989}.  
The rationale for this choice is that HMC schemes with randomized integration times have been 
proven to have accelerated convergence rates for log-concave targets \citep{lu2022explicit} 
compared to HMC schemes with static integration times \citep{chen2019optimal}.
The next question is the choice of the trajectory length distribution. 
No-U-Turn (NUTS) algorithms \citep{hoffman_no-u-turn_2014,betancourt2017conceptual} fill that gap by dynamically determining trajectory length while maintaining invariance with respect to $\bar \pi$. In modern implementations, NUTS is typically combined with a multinomial sampling step to select a point in the generated trajectory as the output of the Hamiltonian approximation kernel \citep[Section A.2.2]{betancourt2017conceptual}.

\paragraph{Step size optimization} We also need to adapt the integrator step size, $\eps$, so that long trajectories can be simulated using as few integration steps as possible without compromising the Hamiltonian error. We use the primal-dual averaging scheme from \citet{hoffman_no-u-turn_2014} targeting an average Metropolis acceptance probability of 80$\%$ (the default value for many HMC packages).

\subsection{Riemannian Manifold methods}\label{sec:riemannian}

When the curvature of the target changes from one region of the state space to the next,
we want our preconditioning metric $M$ to change depending on the location
$\theta$, i.e., $M:=M(\theta)$. By making $M$ position-dependent, we obtain what
is called the Riemannian HMC (RHMC) algorithm \citep{girolami2011riemann}. In this case,
the augmented target becomes
\[
\bar{\pi}_R(\theta,p) := \pi(\theta)\Norm(p|0,M(\theta)) = e^{-H_R(\theta,p)}
\]
where the Hamiltonian is now
\[\label{eq:riemannian-hamiltonian}
H_R(\theta,p) = U(\theta) + \frac{1}{2}\log(|M(\theta)|) + \frac{1}{2}p^\top M(\theta)^{-1}p. 
\]
As with HMC, RHMC is the alternation of two kernels:  (1) a Metropolized kernel with a proposal approximating the Hamiltonian flow (\cref{eq:hamiltonian_dynamics} but with $H$ replaced by $H_R$), and (2) a momentum refreshment. 

\paragraph{First kernel: Metropolized Riemannian Hamiltonian dynamics } The general form of the proposal is the same as standard HMC, 
\[ \rhmc = F \circ \midpoint^T,  \]
but where different integrators are required, denoted $\midpoint$, and described in more detail below. 
In a nutshell, the additional complexity arises because 
of the third term in \cref{eq:riemannian-hamiltonian}, which binds $p$ and $\theta$. As a result, naively using $\hmc$ as a proposal for $\bar \pi_R$ would be neither involutive nor volume preserving \citep[Section~6]{girolami2011riemann}, preventing us from applying \citet{tierney_note_1998}. The implicit alternatives discussed below do have the involutive property and volume preservation \citep[Theorem~3.5]{hairer2006geometric} (up to a controlled numerical error). Therefore, after substitution of $\bar \pi$ by $\bar \pi_R$, the acceptance probability of the Hamiltonian proposal is the same as in \cref{eq:accept}.

\paragraph{Second kernel: position-dependent momentum refreshment } The refreshment step in RHMC mirrors that of HMC, but where the covariance matrix now depends on the position, $p | \theta \sim \Norm(0, M(\theta))$.

Having covered the aspects of RHMC that are similar to basic HMC, we devote the rest of this section to reviewing the aspects that are specific to RHMC.

\paragraph{Integrator choice} As discussed above, the leapfrog integrator is not suitable for RHMC. Fortunately, alternatives that are still involutive and volume preserving have been developed, the most popular being the generalized leapfrog \cite[p. 81--87]{Leimkuhler_Reich_2005,girolami2011riemann} and the implicit midpoint \citep{hairer2006geometric,brofos2021numerical}. We focus on the latter here, denoted $\midpoint$, motivated by its superior performance in a range of target distributions, as reported in \citet{brofos2021evaluating}. 

To define $\midpoint$, let $z$ denote its input, and $z^*$ its output, i.e., $z^* = \midpoint(z)$. Let us introduce the fixed-point function, 
\[ \label{eq:im} g_z(z') := z + \epsilon \bpmat \nabla_p H_R\left(\frac{z'+z}{2}\right)
\\ -\nabla_\theta H_R\left(\frac{z'+z}{2}\right) \epmat, \]
where we write $g_z = g$ for notation simplicity. 
The output value  $z^*$ is defined implicitly using the following fixed-point equation:
\[ z^* = g_z(z^*). \]
This construction is motivated by the fact that symmetry of \cref{eq:im} in $z$ and $z'$ guarantees that $\midpoint$ is involutive. Moreover, from \citet[Theorem~3.5]{hairer2006geometric}, it is also volume preserving. 
In practice, $z^*$ is approximated using iterative numerical methods. 
Under suitable conditions (typically, those needed to apply the Banach fixed-point theorem), we have $g^K(z_0) \to z^*$ as $K\to\infty$, where $z_0 = z$ is customary \citep{hairer2006geometric}. 
This leads to the following fixed-point iterative procedure:
\[
z^{(0)}&=z, \nonumber \\
z^{(k)}&=g\left(z^{(k-1)}\right),\; k=1,\ldots,K, \label{eq:inner-loop}\\
z^* &= z^{(K)},\\
\]
where $K$ is the number of fixed-point iterations. In practice, $K = K(z)$ is determined dynamically by a Cauchy-type criterion. See \cref{app:fp_routine} for a discussion on how $K$ is set and the effect of the target distribution's dimensionality on $K(z)$. 

\paragraph{Metric choice}
In \citet{girolami2011riemann}, $M(\theta)$ is chosen to be the Fisher information matrix of
the target, which is not always available in closed form. A black-box alternative  comes from \citet{betancourt2013general} where $M(\theta)$
is a regularized Hessian matrix of the target negative log density or its diagonal, i.e.,
\[
M(\theta) := r(\nabla^2U(\theta))\quad \text{or}\quad 
M(\theta) := r\left[\diag(\nabla^2U(\theta))\right],
\]
where $r$ is a regularization function to ensure $M(\theta)$ is positive-definite. For computational reasons, we focus on the diagonal variant from now on.
We will also assume that the regularizer 
can be decomposed as 
\[r(A) = \diag(r_1(A_{1, 1}), r_2(A_{2, 2}), \dots, r_d(A_{d, d})), \label{eq:r-decomposition} \]
for some $r_i : \reals \to (0, \infty)$. 
The most popular regularizer is the softabs function \citep{betancourt2013general},
\[
r_i(x) = \text{softabs}_a(x) := x \coth(ax),
\]
where $a$ is a user-defined tuning parameter controlling the smoothness of the softabs near zero. 
The higher $a$ is, the more closely $\text{softabs}_a(x)$ resembles the absolute value function.
The softabs is bounded away from zero and, as such, can limit RHMC's ability to 
explore regions with low curvature (see \cref{app:RHMC_neg_hess} for more details). Therefore, in our numerical implementation, when the target negative log density's diagonal Hessian is already positive, we opt to not apply regularization, i.e., we set $r_i(x) = x$ 
for these cases, and we apply softabs with $a = 5$ otherwise.

\paragraph{Riemannian NUTS} The original NUTS criterion assumes that the metric is position-independent. Therefore, we use the generalized NUTS criterion proposed by \citet{betancourt2013generalizing} 
for all RHMC implementations moving forward. This generalized criterion is also the default 
choice for RHMC in the \verb|AdvancedHMC.jl| package \cite{xu2020advancedhmc}. 
\citet{williams2024geometric} observed that the choice between different variants of these stopping criteria is largely inconsequential for the targets and RHMC variants they considered.

\section{Efficient RHMC for Coordinate Friendly Targets}\label{sec:hess_trick}

\subsection{Coordinate Friendliness}\label{sec:friendliness}

We start by formally defining the \emph{coordinate friendly} property (CF) that we will exploit to compute the Hessian diagonal in compute time $O(d)$. 
Our notion of coordinate friendliness is implied by the one stated in \citet{peng2016coordinate} but is slightly more 
general. Moreover, since \citet{peng2016coordinate} is only concerned about optimization, we provide more context here on how this notion 
is useful within RHMC. 

Let $f : \reals^d \to \reals$ denote the function of interest. In our case, think of $f$ as the energy or negative log density $U$. 
Within each iteration $k$ of the fixed-point solver reviewed in \cref{sec:riemannian}, the CF methods we consider work in two phases. In the {\bf first phase}, we prepare some pre-computed 
\emph{cached storage} based on the current fixed-point iterate $\theta^{(k)}$, $z^{(k)} = (\theta^{(k)}, p^{(k)})$. In the following, we will denote the current fixed-point iterate by $\theta = \theta^{(k)}$, and the associated cached storage by $s = s(\theta)$. Intuitively, the cached storage $s$ is a set consisting of the input $\theta$ and any intermediate values used in evaluating $f$.
In practice, this is done via manipulation of the compute graph, but we defer this discussion to Appendix~\ref{app:implementation} since our method can be understood without knowledge of these implementation details. 
After this first phase is complete, in a {\bf second phase} we  perform forward mode automatic differentiation on $d$ \emph{coordinate functions} $f_{i, s}(x)$, $i \in \{1, \dots, d\}$. The coordinate functions are defined by the property that $s = s(\theta)$ and $\theta = (\theta_1, \dots, \theta_d)$  imply
\[
f_{i, s}(x) = f(\theta_1,\ldots,\theta_{i-1},x,\theta_{i+1},\ldots,\theta_d),
\]
for all $x \in \reals$. To keep the notation concise, we use $f_i(x) = f_{i,s}(x)$ when the dependency on $s$ is clear from the context.

Coordinate friendliness means that the combined computational cost of the two phases described above has the same scaling in $d$ as the computational cost (in flops) of directly computing $f(\theta)$ for $\theta \in \reals^d$. More formally, let $c_0(d)$ denote the cost of storage construction, i.e., evaluation of $s(\cdot)$  (phase 1); $c_i(d)$, $i \in \{1, \dots, d\}$, the cost of evaluation of $f_{i, s}(\cdot)$ (phase 2); and $c(d)$, the cost of direct evaluation of $f(\theta)$. Then we say a method is {\bf coordinate friendly (CF)} if 
\[ \sum_{i = 0}^d c_i = O(c). \]

\bexa \label{eg:cf_plus}
Consider the following function
\[
f(\theta):=\exp\left(\sum_{j=1}^{d}\theta_j\right).
\]
In this case, the cached storage for a given input $\theta^*$ is $s = (\theta^*, v)$ 
where $v = \sum_{j=1}^{d}\theta^*_j$. Note that both $\theta^*$ and $v$ are byproducts of evaluating $f$ at $\theta^*$. Then the coordinate functions at $\theta^*$ can be defined as follows
\[
f_{i,s}(x) = \exp(x - \textcolor{blue}{\theta^*_i} + \textcolor{blue}{v}).
\]
Note that the terms in blue all come from the cached storage $s$.
Finally, we can see that evaluating $f$ costs $O(d)$ (first phase) while in the second phase, the cost of evaluating $f_{i,s}(x)$ for one given $x$ costs $O(1)$. This makes $f$ a CF function.
\eexa

We show in Appendix~\ref{app:implementation} that many target distributions commonly used in statistics can be implemented in a coordinate friendly manner. Specifically, we show this is the case for all sparse graphical models as well as generalized linear models.

\subsection{CF Reverse over Forward Method}\label{sec:main-method}

Now we introduce our linear-scaling implementation of RHMC for a CF target $\pi$, 
which we call compute graph RHMC (CG-RHMC). Our implementation involves utilizing
different modes of automatic differentiation (AD) at different stages of the algorithm.
To distinguish between these modes of AD, we denote \forwardAD\ the forward mode AD 
operation, \reverseAD\ the reverse mode AD operation, \JacobianAD\ and \HessianAD\ 
the standard way AD libraries evaluate the Jacobian matrix and Hessian matrix, respectively
(as a convention,  
we use \texttt{monospace fonts} to emphasize when a function uses a specific numerical implementation).
These operations take in a function and its input value
as arguments, then return the gradient, Jacobian or Hessian at that input value.
For scalar input functions where higher-order derivatives can be obtained through 
repeated application of forward mode AD, we denote 
\[
\forwardAD^2(f,x):= \forwardAD(\forwardAD(f,\cdot),x),
\]
where $f$ is a $\reals\to\reals$ function and $x\in\reals$ the input value.
For more background on these AD functions, see \cref{app:AD_modes}. 

We will start by discussing the computation of the Hamiltonian, and then turn to its gradient.

\subsubsection{Computation of \texorpdfstring{$H_R$}{HR} for CF Targets}

Computation of $H_R$ requires evaluation of the metric
$M(\theta)$, which recall that in our setup we are focusing on the case where 
$M(\theta) :=r(\diag(\nabla^2 U))$, where $r$ denotes an optional regularization to ensure 
the metric is positive-definite, which we assume can be written as in Equation~(\ref{eq:r-decomposition}).  

Since $\pi$ is assumed to be CF, so is $U$, and so let $U_{i,s}$ for $i \in \{1, \dots, d\}$ denote its coordinate functions. 
We denote by $\CGM(s, \theta)$ a CF implementation of $M(\theta)$ utilizing the 
cache storage $s = s(\theta)$. It is constructed as follows:  
\[
\CGM(s, \theta) &:= \diag\left( \CGM_1(s, \theta), \dots, \CGM_d(s, \theta)\right),  \\
\CGM_i(s, \theta) &:= \forwardAD^2(r_i \circ U_{i, s}(\cdot), \theta). 
\]
Here the notation $r_i \circ U_{i, s}(\cdot)$ can be understood as building 
a temporary anonymous function, defined by applying $U_{i, s}$ followed by $r_i$. 
That anonymous function is then fed into the forward differentiation routine. 
As our notation makes explicit, even though $s$ is obtained upstream by computing 
$s = s(\theta)$, for the point of view of the \forwardAD\ call it is viewed as a constant. 

Based on \citet[Chapter~4]{griewank2008evaluating}, 
the cost of applying \forwardAD\ to a function $f:\reals\to\reals$ is at most a constant 
multiple of its evaluation cost. Since evaluating $U_{i, s}(\cdot)$ has a runtime cost of $c_i$,
$\forwardAD(r_i \circ U_{i, s}(\cdot), \theta)$ costs $O(c_i)$ to evaluate and so does 
$\forwardAD^2(r_i \circ U_{i, s}(\cdot), \theta)$.
Therefore, evaluating the \CGM\ function costs $\sum_{i=1}^{d} c_i$ 
which is $O(c)$ thanks to the CF assumption. 

Next, we turn to the definition of a function \Hfunc\ that takes in a diagonal metric $\M$ 
and a phase point $z = (\theta, p)$, and returns the Riemannian Hamiltonian 
$H_R(z)$ (Algorithm~\ref{algo:H}). More precisely, here $\M$ is any  
implementation of a function computing the metric, the key example being $\CGM(s, \cdot)$. 

\begin{algorithm}
\caption{Hamiltonian evaluation for generic RHMC with diagonal mass matrix}\label{algo:H}
\begin{algorithmic}
\Require \\
diagonal metric evaluation function \M \\
phase state $z = (\theta, p)$
\Ensure Riemannian Hamiltonian $H_R(z)$
\Function{\Hfunc}{$\M, z$}
\State $M \gets \M(\theta)$ \Comment{$O(c_M)$}
\State \textbf{return}  $U(\theta) +\frac{1}{2}\sum_{i=1}^{d} \left(\log M_{i,i} + \frac{p_i^2}{M_{i,i}}\right)$ \Comment{$O(d)$}
\EndFunction
\end{algorithmic}
\end{algorithm}

We keep the input \M\ generic in the definition of \Hfunc\ to make it easier to compare the efficiency of our implementation with that of previous work in the next section. Additionally, we use the notation $c_M$ for the cost of evaluation of \M, for example, when $\M = \CGM$, $c_M = O(d)$. 

We denote our proposed forward AD based implementation by:
\[
\CGH(s, z) := \Hfunc(\CGM(s, \cdot), z).
\]

\subsubsection{Computation of \texorpdfstring{$\nabla H_R$}{nabla HR} for CF targets}

Having established how to compute $H_R$, we turn to its gradient. That computation is critical as it occurs in the inner-most loop of RHMC, namely in the fixed-point update $g$ defined in Equation~(\ref{eq:inner-loop}). 
The cost of evaluating $g$ boils down to the evaluation of $\nabla H_R$. 

To efficiently compute $\nabla H_R$, we apply again automatic differentiation, 
but this time in its reverse mode:
\[
\CGGradH(\theta, p) := \reverseAD(\CGH(s(\theta), \cdot), (\theta, p)).
\]
Recall that the notation $\CGH(s(\theta), \cdot)$ corresponds to an anonymous function, 
where only that anonymous function's argument will be the differentiated argument 
in the \reverseAD\ call. Hence, as before, $s = s(\theta)$ is treated as a constant. 

Based on the cost breakdown in \cref{algo:H} and the $O(c)$ cost of \CGM, 
the cost of \CGH\ is $O(c)$, so applying \reverseAD\ on top of it will also 
cost $O(c)$ (\citep[Chapter~4]{griewank2008evaluating}). 

To compute a single implicit step, we then use Algorithms 1 and 3 in  \citet{brofos2021evaluating} where 
we plug in our gradient computation function \CGGradH\ for computing Equation~(10) in that paper.

\subsection{Comparison to CF-unaware methods}\label{sec:comparison}

In this section, we compare the computational cost of the method described in the previous section 
to other methods in the literature that use a diagonal metric, but do not exploit a CF structure.
We call these methods ``CF-unaware.'' 
We emphasize that all these methods lead to the same Hamiltonian trajectories up to numerical precision, a property that we use to validate our implementation (see Figure~\ref{fig:trajectory}). 

Building on the notation from the previous section, the most natural CF-unaware approach is 
to change the metric function \M\ that we input into Algorithm~\ref{algo:H}: instead of using 
the CF-informed version, $\M = \CGM$, we can use a CF-unaware version, $\M = \BlindM$, where:
\[ \BlindM(\theta) = \diag(\HessianAD(U, \theta)). \]
This leads to a cost for metric calculation proportional to $d$ times the evaluation cost of $U$, i.e., $c_M = O(d c)$ \citep[Section 6]{Martens2012Estimating}. Tracing the same reasoning as in the last section, we then construct:
\[
\BlindH(s, z) &:= \Hfunc(\BlindM, z), \\
\BlindGradH(z) &:= \reverseAD(\BlindH, z).
\]
Again, by the same argument as last section, but with $c_M = O(cd)$ instead of $c_M = O(c)$, we obtain that \BlindGradH\ has a running time of $O(cd)$. 

Surprisingly, when numerically benchmarking the runtime scaling of an implementation of RHMC in the popular \texttt{AdvancedHMC.jl} package \citep{xu2020advancedhmc}, we found that the observed runtime scaling was higher than the one predicted by the above analysis of \BlindGradH\ (see Figure~\ref{fig:timing}, in the benchmark model, $c = O(d)$, so we expected a cost scaling of $O(d^2)$ for the CF-unaware method but observed $O(d^3)$). In the rest of this section, we briefly explain the cause of this extra cost. We call this third approach, the ``suboptimal method'' since it is strictly dominated by the other ones. 

The suboptimal cost arises when computing $\nabla H_R$ and stems from the use of the following formulas, derived in \citet{girolami2011riemann}
\[
\nabla_{p_i} H_R(\theta,p) &= \{M(\theta)^{-1}p\}_i \label{eq:dH_p}\\
\nabla_{\theta_i} H_R(\theta,p) &= \nabla_{\theta_i}U(\theta) + 
\frac{1}{2}\left\{\text{tr}\left(M(\theta)^{-1}\nabla_{\theta_i}M(\theta)\right) +
p^\top M(\theta)^{-1} \nabla_{\theta_i}M(\theta) M(\theta)^{-1}p \right\}\\
&= \nabla_{\theta_i}U(\theta) + 
\frac{1}{2} \sum_{j=1}^{d} \left\{ \frac{\nabla_{\theta_i}M_{j,j}(\theta)}{M_{j,j}(\theta)} +
\nabla_{\theta_i}M_{j,j}(\theta) (\nabla_{p_i} H_R(\theta,p))^2 \right\} \label{eq:dH_theta}.
\]
As we detail in Algorithm~\ref{algo:suboptimal}, computing $\nabla H_R$ via evaluation of each term in \cref{eq:dH_theta} yields a cost of $O(d c_M)$, which in the CF-unaware case is $O(c d^2)$. 
This bottleneck is because computing each term in \cref{eq:dH_theta} is equivalent to the computation of $\JacobianAD$ for $\M$.
\footnote{We note that this bottleneck shows up not only in \texttt{AdvancedHMC.jl} but also in many other code bases that implement RHMC.
See for example: \href{ https://github.com/stan-dev/stan/blob/3884f60c54261c9b3a51a6147a6d3f6d2a7060c6/src/stan/mcmc/hmc/hamiltonians/softabs_metric.hpp\#L60 }{\texttt{github.com/stan-dev/stan}}, 
\href{https://github.com/rsantet/RMHMC/blob/dd731550dd253403ab3e8cad39047d091025ff51/main_nd.jl\#L39}{\texttt{github.com/rsantet/RMHMC}},
\href{https://github.com/JamesBrofos/Thresholds-in-Hamiltonian-Monte-Carlo/blob/7ee1b530db0eb536666dbc872fbf8200e53dd49b/hmc/applications/stochastic_volatility/stochastic_volatility.py\#L350}{\texttt{github.com/JamesBrofos/Thresholds-in-Hamiltonian-Monte-Carlo}}
}
In contrast, using \reverseAD\ on $H_R$ is equivalent to viewing the sum in \cref{eq:dH_theta}
as an element in a vector-Jacobian product, which evaluates $\nabla H_R$ in $O(c_M)$ time. 
Note, however, that this efficient computation of $\nabla_\theta H_R$
still costs $O(c d)$, which is quadratic, in the CF-unaware case.
\section{Experiments}\label{sec:experiment}

\subsection{Experimental Setup Generalities and Description of Test Problems}\label{sec:ex-setup}

\if1\anon
{
The source code for these experiments can be found at \url{https://github.com/UBC-Stat-ML/CG-RHMC-mev}.
All methods are implemented in Julia and use components of the package \verb|AdvancedHMC.jl| v0.7.1 \citep{xu2020advancedhmc} whenever possible, making wall clock time comparisons sensible. 
To perform the reverse-over-forward automatic differentiation, we use the package \verb|Enzyme.jl| v0.13.129 \citep{NEURIPS2020_9332c513}. 
} \fi

\if0\anon
{
The source code for these experiments can be found at [anonymized URL] (also included in the supplement zip file).
All methods are implemented in Julia and use components of the package \verb|AdvancedHMC.jl| v0.7.1 \citep{xu2020advancedhmc} whenever possible, making wall clock time comparisons sensible. 
To perform the reverse-over-forward automatic differentiation, we use the package \verb|Enzyme.jl| v0.13.129 \citep{NEURIPS2020_9332c513}. 
} \fi

For the funnel target, we consider the following
distribution with scale parameter $\beta>0$ given by
\[
X_1&\sim N(0,3^2)\\
X_i|X_1=x_1 &\sim N(0,e^{x_1/\beta})\quad i=2,3,\ldots,100
\]
Here we set $\beta = 0.5$ which makes this target significantly harder than the 
original funnel target in \citep[Section~8]{neal_slice_2003} where $\beta = 2$.

We also consider the following horseshoe logistic regression model:
\[
y_j &\sim \Bern(\logistic\left(x_j^\top \theta\right)), & \forall\ j\in\{1,\ldots,n\}\\
\theta_1 &\sim \StudentT (3,0,1) \\
\theta_i &\sim \Norm(0,(\lambda_i^*)^2(\tau^*)^2), & \forall\ i\in\{2,\ldots,d\} \\
\lambda_i^* &\sim \Cauchy^+(0,1), & \forall\ i\in\{2,\ldots,d\} \\
\tau^* &\sim \Cauchy^+(0,1),
\]
where $x_j\in \reals^d$ are the covariates, $y_1,\ldots,y_n\in\{0,1\}$ are the 
responses and $n$ the sample size. Additionally, $\StudentT(\nu,\mu,\sigma)$ denotes a 
t-distribution with $\nu$ degrees of freedom, location $\mu$, scale $\sigma$ and 
$\Cauchy^+(0,1)$ denotes a standard half-Cauchy distribution.
Since $\lambda_i^*$ and $\tau^*$ make the prior undefined at zero, we transform
the model to the unconstrained space by setting $\lambda_i := \log((\lambda_i^*)^2)$
and $\tau := \log((\tau^*)^2)$.
Both the funnel and the horseshoe logistic regression problems are CF.

\subsection{Validation of the Implementation}\label{sec:correct}

We outline here some of the checks we used to ensure our implementation is correct, which in our context means that it outputs exactly the same trajectory as previous RHMC methods (up to numerical precision), the only change being the running time. 

\begin{figure}[t]
	\centering
	\begin{subfigure}{0.48\textwidth}
		\centering
		\includegraphics[width=0.95\textwidth]{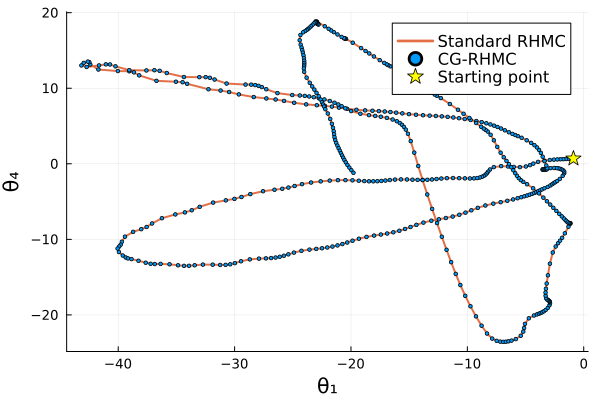}
	\end{subfigure}
	\begin{subfigure}{0.48\textwidth}
		\centering
		\includegraphics[width=0.95\textwidth]{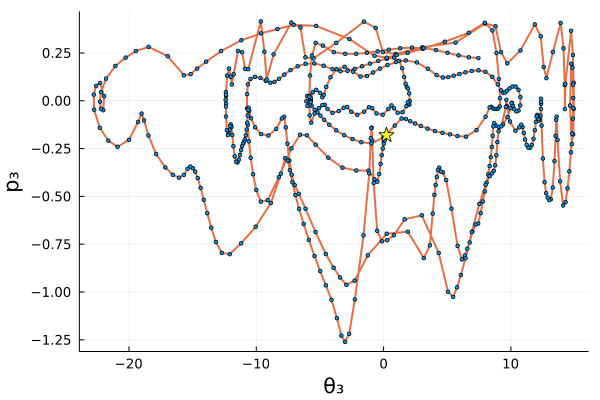}
	\end{subfigure}
	\caption{Trajectory comparison between CG-RHMC and standard RHMC for an 8-dimensional logistic
		regression problem. \textbf{(Left)} position trajectory of $(\theta_1,\theta_4)$, \textbf{(right)} phase space trajectory of $(\theta_3,p_3)$.}
	\label{fig:trajectory}
\end{figure}

As a first verification, we compared two implementations of RHMC: one third-party implementation based on \cref{eq:dH_theta} provided in the \verb|AdvancedHMC.jl| package, and ours, based on \cref{sec:main-method}.
We used the same midpoint integrator for both, with the only 
change being the functions to compute the Riemannian 
Hamiltonian and its gradient, which is the focus of our work. 
Denote the two integrator implementations by $\midpoint^{\text{baseline}}$ and $\midpoint^{\text{CF}}$. 
Here, we used a logistic regression model with $d = 8$, 
and generated an initial point $z_0 = (\theta_0, p_0)$ according to a standard normal for 
$\theta_0$ and to the RHMC momentum distribution for $p_0$. 
For the two integrators $i \in \{\text{baseline}, \text{CF}\}$, we build a trajectory $z_t^{(i)}$, both initialized at $z_0^{(i)} = z$ and constructed in the usual way, $z_t^{(i)} = \midpoint^{(i)}(z_{t-1}^{(i)})$ for $t > 0$. We show two representative bivariate marginals of the resulting trajectories in \cref{fig:trajectory}, one capturing two parameters, and the other, a parameter with its dual momentum. As expected, the two trajectories are identical up to numerical precision. 

Next, to validate our midpoint integration implementation, we compared it to a standard non-reversible parallel tempering (NRPT) \citep{syed2022non} baseline using Slice sampling with doubling \citep{neal_slice_2003} as \emph{explorers} (i.e., local MCMC kernels). This is to ensure that the NRPT baseline does not use any position-dependent pre-conditioning. Since NRPT relies on Metropolized swaps between tempered chains to achieve a high quality posterior approximation, the fact that these two very different posterior approximation methods yield very similar results (see Figures~\ref{fig:horseshoe_marginals}, \ref{fig:horseshoe_ks}, \ref{fig:horseshoe_pairplot}) supports that they target the correct distribution.

\subsection{Compute graph RHMC is faster than baseline implementations}\label{sec:faster}

\begin{figure}[t]
  \centering
  \includegraphics[width=0.7\textwidth]{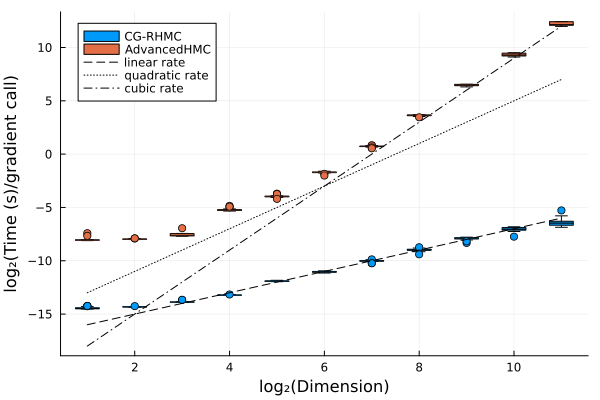}
  \caption{Wall-clock time (in seconds) per Hamiltonian gradient call versus dimension for 
  different RHMC implementations on synthetic logistic regression datasets of increasing dimensionality 
  (log-log scale, lower is better). }
  \label{fig:timing}
\end{figure}

We study the running time of two implementations of $\nabla H_R$: our method, labelled \texttt{CG-RHMC}, and the implementation from the package
\verb|AdvancedHMC.jl| v0.7.1 \citep{xu2020advancedhmc} which uses the package
\verb|Zygote.jl| v0.6.77 \citep{Zygote.jl-2018} as the AD wrapper for their metric computations. 
For this experiment, we consider a sequence of synthetic 
logistic regression datasets with increasing dimension $d \in \{2^1,\ldots,2^{11}\}$.
Each parameter has a Gaussian prior with standard deviation 10.
This prior makes the target distribution log-concave and its negative log density
Hessian positive-definite. Therefore, no metric regularization is applied.
For each implementation, we run 12 gradient calls of $H_R$ and
use the average time per gradient call 
as a representation of the computational complexity.
The results, presented in \cref{fig:timing}, support that CG-RHMC  
achieves an $O(d)$ scaling (the same as density evaluation complexity), 
while the standard implementation of RHMC scales as $O(d^3)$.
The reason for the $O(d^3)$ scaling despite $M(\theta)$ being diagonal is due to
the computational pitfall mentioned at the end of \cref{sec:hess_trick}.

\subsection{Comparison between position-dependent and position-independent metrics}

There is plenty of numerical evidence in the literature supporting that for the same number of integrator steps, RHMC performs better than HMC \citep{girolami2011riemann,kailas2026hierarchical}. However, because of the higher compute cost per integrator step, in previous studies it was often observed that the ranking between RHMC and HMC is often reversed when looking instead at sample quality normalized by time, which is ultimately the quantity of interest. 

In previous work the higher cost per integration step came from two factors: the higher cost per gradient call, and the need for implicit solvers. 
By exploiting the CF structure in the target, we have addressed the first, but our method still requires an implicit solver. 

The goal of this section is to investigate whether that improvement is sufficient to make RHMC competitive with state-of-the-art methods, including HMC and NUTS methods that use a fixed but adapted diagonal mass matrix. Our findings support that indeed our computational method makes RHMC attractive on a time normalized basis.

\subsubsection{Funnel target}

In this next experiment, we run CG-RHMC, CG-RNUTS, standard RHMC, standard RNUTS, HMC and NUTS on 
a scaled 100-dimensional funnel distribution and compare their rates of convergence via the 
2-Wasserstein distance of the first coordinate marginal. 

For non-Riemannian methods, we use both unit and diagonal metrics and tune the
diagonal metrics using sample variance. There are other ways 
to tune non-Riemannian metrics \cite{hird2023precondition,hird2026high}, but we opt to use the most widely used method here.
Note also that the negative log density of the funnel has positive Hessian diagonal, 
so no metric regularization is used.
For static trajectory length methods, we use $l=20$ integration steps for each trajectory. 
Each algorithm follows the procedure below: (1) perform a pilot run for 20000 iterations with the 
first 10000 iterations used to adapt the step size (and metric) to achieve $80\%$ acceptance rate
(2) use the adapted step size to run chains with no adaptations for a fixed amount 
of time (20 seconds) across 1000 different seeds
(3) use samples across different seeds to estimate the 2-Wasserstein distance between 
the first coordinate marginal of the chains and that of the target distribution.
The 2-Wasserstein distance in the above procedure is approximated using
the formula for 2-Wasserstein distance between two 1-dimensional 
ordered samples $x_{1:B},y_{1:B}$
\[
W_2(\hat{\mu}_B,\hat{\nu}_B) := \sqrt{\frac{1}{B}\sum_{b=1}^{B}(x_b-y_b)^2}
\]
where $\hat{\mu}_B,\hat{\nu}_B$ are empirical distributions of $x_{1:B},y_{1:B}$, respectively.
In the procedure above, $x_{1:B}$ are the ordered statistics first elements of the chain states at a 
fixed iteration for a given algorithm and $y_{1:B}$ are the ordered statistics from $B$ independent draws from the true first marginal distribution of the funnel.

\begin{figure}[t]
  \centering
  \includegraphics[width=0.95\textwidth]{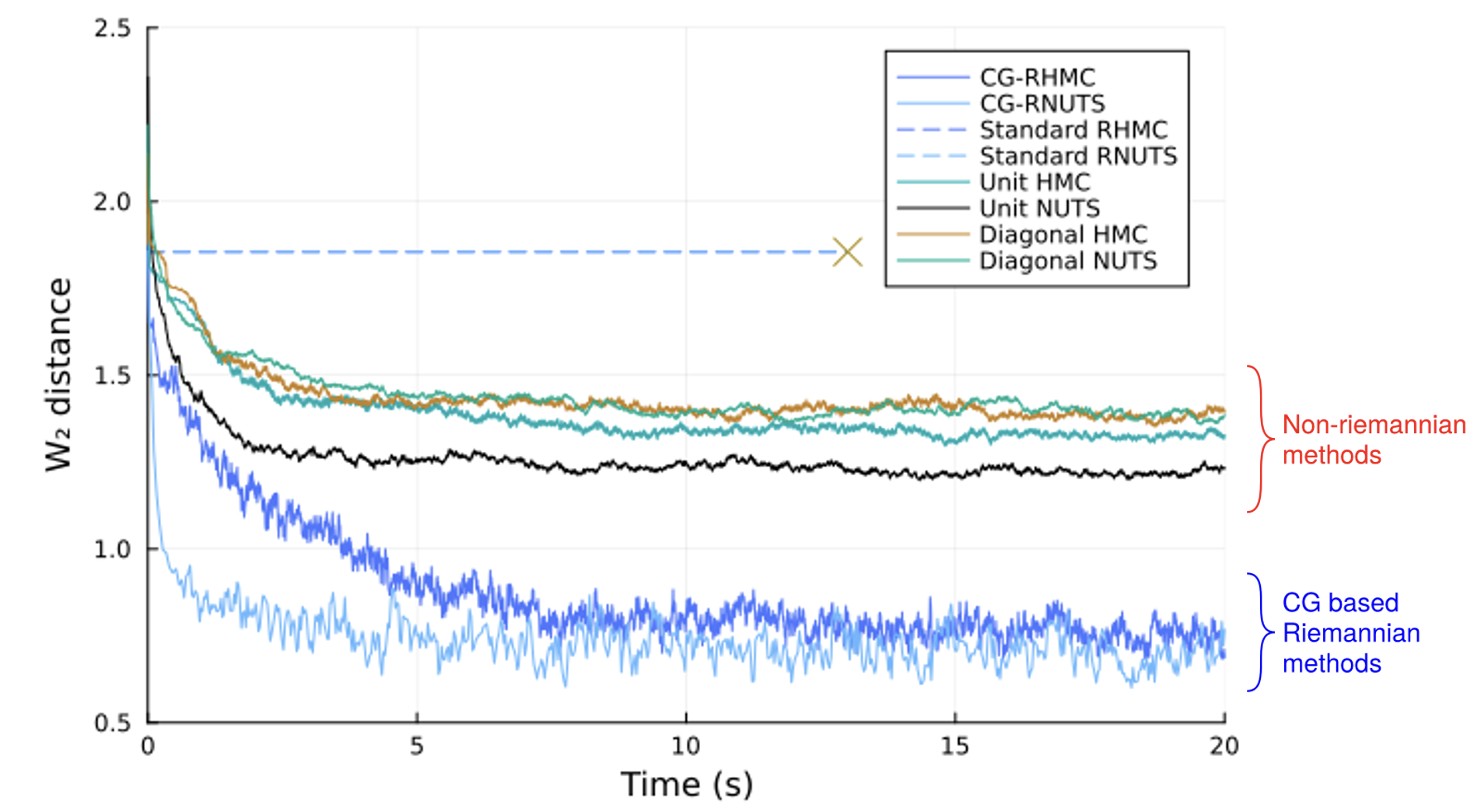}
  \caption{First coordinate 2-Wasserstein distance convergence over time 
  for various HMC and RHMC variants on a 100-dimensional funnel target
  (log-log scale, lower is better). The blueish lines are Riemannian methods and the black
  line is the best non-Riemannian method. Dotted lines are standard implementations of RHMC and
  RNUTS which did not finish obtaining one sample in the allotted time 
  (indicated by the \textcolor{red}{$\times$} marking). }
  \label{fig:funnel}
\end{figure}

The results are illustrated by \cref{fig:funnel}. Based on this, we can see that our implementation
has the best 2-Wasserstein distance overall and demonstrates the potential of RHMC in high-dimensional settings.
In contrast, standard RHMC and RNUTS did not even finish obtaining one sample for this 
100-dimensional target, which is a relatively small high-dimensional target.

Since the 2-Wasserstein distances are estimated from a finite set of samples (1000 independent chains), we are not able to measure very small distances, explaining the apparent lack of convergence to zero. In reality, all methods considered are $\pi$-invariant.

\subsubsection{Real data experiment}\label{sec:realdata_experiment}

In this experiment, we compare the convergence of NUTS and CG-RNUTS to stationarity for 
a logistic regression model with the horseshoe prior \citep{carvalho2009horseshoe} on
a real dataset. In order to produce a reference posterior, we run a long chain using the non-reversible parallel tempering (NRPT)
algorithm \cite{syed2022non} and compare samples from NUTS and CG-RNUTS
to that of NRPT. For CG-RNUTS, we run a chain for $64\times 10^3$ iterations with half of them used for 
adapting the step size and then thrown away. For NUTS, we run a chain for $512\times 10^3$ iterations 
with half of them used for adapting the step size, metric and then thrown away. The difference
in iterations is to ensure both samplers are given roughly the same sampling time budget
(CG-RNUTS and NUTS took $3.8\times 10^3$ and $6.3\times 10^3$ seconds in total, respectively). 
For NRPT, we use the implementation from the \verb|Pigeons.jl| package \citep{surjanovic2025pigeons} 
and run a $2^{16}\approx 64\times 10^3$ iteration, 10-replica chain (with each iteration comprised of one odd or even swap, and 3 sweeps of local explorer) with half of them used for adaptation and then thrown away. For horseshoe logistic regression targets, we use the horseshoe prior for the reference chain of NRPT.
See \cref{app:NRPT_correctness} for diagnostics supporting that the NRPT chains have reached stationarity. 

In this section, we use a version of the prostate cancer dataset from \citet{singh2002gene} where the data dimension has been reduced from 5966 to 50 covariates by \citet{biron2024automala} (see \cref{app:exp} for more details). This is done to ensure the NRPT chain, which in this experiment is based on a $O(d^2)$ naive slice sampling algorithm, will converge to stationarity in a reasonable amount of time.
Additionally, we subsample 10 observations from the dataset and apply the horseshoe
logistic regression model on the subsampled data. By using a small number of observations,
the posterior can exhibit varying geometries as observed in \cref{fig:horseshoe_pairplot}, 
increasing the difficulty of the problem. For a comparison to the posterior with no subsampling,
see the bivariate marginals in \cref{fig:horseshoe_pairplot_full}.

\begin{figure}[t]
  \centering
  \begin{subfigure}{0.50\textwidth}
    \centering
    \includegraphics[width=0.95\textwidth]{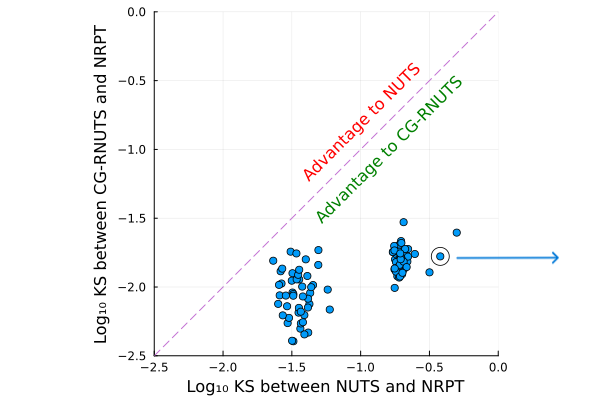}
  \end{subfigure}
  \begin{subfigure}{0.45\textwidth}
    \centering
    \raisebox{0.5cm}{\includegraphics[width=0.95\textwidth]{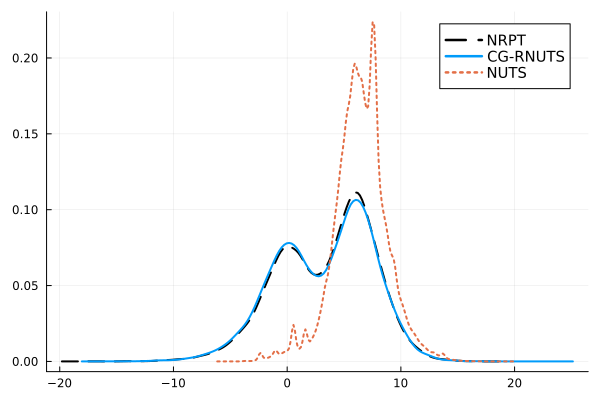}}
  \end{subfigure}
  \caption{\textbf{(Left)} KS test statistics of each parameter between the NRPT chain (reference posterior) 
  and the two competing algorithms NUTS and CG-RNUTS in log-log scale (lower is better) 
  for the horseshoe logistic regression problem on the prostate cancer data with similar time
  budgets where each point represents a parameter. \textbf{(Right)} Marginal density comparison
  for the circled point.}
  \label{fig:horseshoe_ks}
\end{figure}

In the left facet of \cref{fig:horseshoe_ks}, we plot the two-sample Kolmogorov-Smirnov (KS) test
statistics between samples from the competing algorithms and the sample from the NRPT chain 
for each model parameter, i.e. we compare the marginal distribution of each parameter from 
NUTS and CG-RNUTS to the reference posterior. In the right plot of \cref{fig:horseshoe_ks},
we see the disparity between the marginals of NUTS and the other two methods: the NUTS chain
was unable to explore the left tail of the target. Not only that, we can also see how the 
NUTS chain differs from the NRPT and CG-RNUTS chains through \cref{fig:horseshoe_marginals}
and the bivariate marginals in \cref{fig:horseshoe_pairplot}. It is clear that, given the same 
\emph{reasonable} time budget, CG-RNUTS is the better choice compared to NUTS for this kind of model.
We must emphasize that standard RNUTS can achieve the same sample quality but at a much steeper
time budget (200 times that of CG-RNUTS for this target).

\subsubsection{Panel of datasets}\label{sec:panel_dataset}

\begin{figure}[t]
  \centering
  \includegraphics[width=0.95\textwidth]{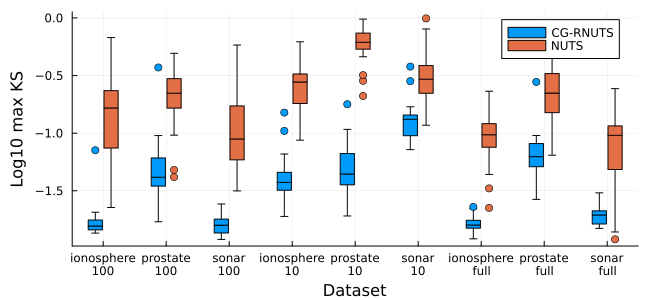}
  \caption{Box plot of maximum marginal KS statistics of CG-RNUTS and NUTS 
  compared to the NRPT reference posterior (lower is better) for horseshoe logistic regression 
  targets with various datasets and data sizes.}
  \label{fig:panel_max}
\end{figure}

We now repeat the experiment in \cref{sec:realdata_experiment} with different datasets and data sizes.
Specifically, we consider an ionosphere radar returns dataset \citep{sigillito1989ionosphere}, a sonar signal dataset \citep{sejnowski1988connectionist}
and the prostate cancer data from the last experiment. For each dataset, we run CG-RNUTS and NUTS on
three different data size settings: the full dataset, a random 10 observation subsample and 
a random 100 observation subsample--for a total of 9 different targets. For each target, the 
experiment in \cref{sec:realdata_experiment} is repeated 20 times with different seeds. The maximum
marginal KS statistics of CG-RNUTS and NUTS compared to the NRPT reference posterior are shown in \cref{fig:panel_max}
(see \cref{fig:panel_stats} for results where the KS statistics are summarized by median, mean and minimum).
As expected, CG-RNUTS chains produce results closer to the reference posterior than NUTS with similar time budgets
(see \cref{fig:panel_time} for time budget comparison).

\section{Disclosure statement}\label{disclosure-statement}

The authors report there are no competing interests to declare.

\section{Data Availability Statement}\label{data-availability-statement}

\if1\anon
{
Deidentified data have been made available at the following URL: \url{https://github.com/UBC-Stat-ML/CG-RHMC-mev}.
} \fi

\if0\anon
{
Deidentified data have been made available in the supplement zip file.
} \fi

\bibliography{main.bib}

\newpage

\phantomsection\label{supplementary-material}
\bigskip

\begin{center}

{\large\bf SUPPLEMENTARY MATERIAL}

\end{center}

\appendix
\section{Implementation of coordinate friendly methods via compute graphs}\label{app:implementation}

\subsection{Compute graph}

The compute graph is a data structure popularized by the automatic differentiation literature, in which it plays a central role in reverse-mode automatic differentiation \citep[p.19]{griewank2008evaluating}. 
Here we use it for an additional, distinct purpose, for coordinate friendly evaluation, described in \citet[Section 3]{luu2025gibbs}.
In both cases, a key property of the compute graph is that there are now popular software tools that can automatically extract it from arbitrary code, for example using software engineering techniques such as tracing and source code transformation \citep{griewank2008evaluating}. 

The \textit{compute graph} of a distribution  encodes the order of operations 
in the computation of its density.
While the compute graph superficially looks similar to directed graphical models, the two are distinct concepts. 
In the case of the compute graph, only source nodes correspond to random variables while each non-root node represents a function that takes the values of its 
parent nodes (nodes that point to it) and performs a primitive operation (e.g., addition, multiplication, etc).
To perform a forward pass or evaluation of the compute graph, we simply evaluate each
function node of the graph using its parent nodes as input and store the output in the node
until we reach the sink node. 
In our setup, there is a single sink node, corresponding to the negative log density $U(\theta)$.

\subsection{Coordinate friendly compute graphs}

Compute graphs provide a powerful tool to automate the implementation of the abstract notion of coordinate friendliness 
defined in Section~\ref{sec:hess_trick}. We explain in this section how this is done in practice. 

Recall the two \emph{phases} defined in \cref{sec:friendliness}. In the following, we describe how they are implemented in the context of a compute graph. 

In the first phase, the computation of the cached storage, $s(\theta)$, corresponds to forward evaluation of the compute graph, but where the value of each node in the graph is inserted in a cache (which can be conceptualized as a dictionary where keys are nodes in the compute graph, and values are the output of the node). This cache is the object $s$ returned by $s(\theta)$. 

Next, for the second phase, we show how the evaluation of each coordinate function $f_{i,s}(\cdot)$ can be done efficiently in \emph{CF compute graphs}. 
To do so, we examine the subgraph of the compute graph obtained by keeping only the nodes having the compute graph's input node $\theta_i$ as an ancestor. 
From here on, we refer to this type of subgraph as the \textit{$i^{th}$ coordinate subgraph}, and the compute graph of $f$ as the full graph. 
Thanks to the storage of $s$ from the first phase, notice that evaluation of $f_{i,s}(\cdot)$ only involves computations 
at the nodes in the coordinate subgraph
(see \cref{eg:glm,eg:AR} for a visualization of coordinate subgraphs).

Moreover, when one of the nodes is a sum over a large number of incoming edges, it is possible 
to update it at a cost independent of the number of outdated incoming edges, as illustrated in the following example.

\bexa \label{eg:cg_plus}
Consider the function
\[
f(\theta):=\sum_{j=1}^{d}\theta_j.
\]
The compute graph for this function includes the input nodes $\theta_1, \dots, \theta_d$, all of which point to a summation node, ``$+$''.
After a forward pass, the full graph stores the $\theta$ values in the $\theta$ nodes and
the value of $f(\theta)$ in the $+$ node. 
In this case, the $i^{th}$ coordinate subgraph is just the $\theta_i$ node pointing to the
$+$ node. The function $f_i$ in the $+$ node of the coordinate subgraph can be defined in 
different ways as follows
\[
f_i(x) &= x + \sum_{j\neq i}\theta_j, \\
 &= x - \theta_i + \sum_{j=1}^d\theta_j, \\
 &= x - \theta_i + \underbrace{f(\theta)}_\text{available from the stored cache $s$.} \label{eq:def3}
\]
We can see that the forward pass on the full graph 
costs $O(d)$ while the forward pass on the coordinate subgraph costs $O(1)$
when \cref{eq:def3} is used in its evaluation.
\eexa

\cref{eg:cg_plus} illustrates the type of function definition we want in coordinate subgraphs 
for efficient evaluation. We can generalize that example

\bexa
Consider a compute graph corresponding to a commutative group operator node, $\bullet$, 
with $d$ parent nodes, $x:=(x_1,\ldots,x_d)$. Then we can define the $\bullet$ node
in the $i^{th}$ coordinate subgraphs as
\[
f_i(x'_i) := x'_i \bullet (-x_i) \bullet f(x)
\]
where $f(x)$ is the value stored in the $\bullet$ node of the full graph and $-x_i$ is
the inverse element of the group induced by the $\bullet$ operator.

If the group operation is non-commutative, a single edge update can still be performed 
in $O(\log d)$, by using associativity to express the product as a tree of depth $O(\log d)$.
\eexa

Using these tools, the class of posterior distributions that admit a CF compute graph covers many models including generalized linear models (GLMs), additive
models, and Bayesian hierarchical models.
We provide examples below, showing at the same time that the compute graph representation 
can lead to coordinate friendly implementation in cases where the graphical model would fail to be 
coordinate friendly.

\bexa \label{eg:glm}
Consider a generalized linear model with independent priors $\pi_0$. 
The negative log joint density, $U$, for such target has the following form 
\[
p_j(\theta) &= \logistic\left(x_j^\top \theta\right)\\
U(\theta) &= \sum_{j=1}^{n}\underbrace{\left[y_j\log p_j(\theta)+(1-y_j)\log(1-p_j(\theta))\right]}_{l_j(\theta)} 
\underbrace{- \sum_{i=1}^{d}\log \pi_0(\theta_i)}_{U_0(\theta)}
\]
where $x_j\in \reals^d$ are the covariates, $y=(y_1,\ldots,y_n)\in\{0,1\}^n$ are the 
responses and $n$ the sample size. For a fixed coordinate $k$, we are interested in 
computing $U_k(\theta_k')$ given $\theta$ and the cached storage $s$ obtained from a full forward evaluation of $U(\theta)$.
For coordinate subgraph evaluation, we look at the evaluation and cost breakdown of 
the compute graph of $U$ and that of its $k^{th}$ coordinate subgraph side by side:

	\textbf{Full graph evaluation}
	\[
	&\text{Set }L\leftarrow 0 \\
	&\textbf{For } j=1,\ldots,n & \triangleright\ O(dn)\\
	&v_j \leftarrow x_j^\top \theta & \triangleright\ O(d)\\
	&p_j \leftarrow \logistic(v_j) & \triangleright\ O(1)\\
	&l_j \leftarrow y_j\log p_j\\
	&\qquad +(1-y_j)\log(1-p_j) & \triangleright\ O(1)\\
	&L \leftarrow L + l_j & \triangleright\ O(1)\\
	&\textbf{end}\\
	&\text{Set }U_0\leftarrow 0 \\
	&\textbf{For } i=1,\ldots,d & \triangleright\ O(d)\\
	&U_0 \leftarrow U_0 - \log \pi_0(\theta_i) & \triangleright\ O(1)\\
	&\textbf{end}\\
	&U\leftarrow L + U_0 & \triangleright\ O(1)\\
	&\textbf{Return } U
	\]

	\textbf{Coordinate subgraph evaluation}
	\[
	&\text{Set }L_k\leftarrow 0 \\
	&\textbf{For } j=1,\ldots,n & \triangleright\ O(n)\\
	&v_j' \leftarrow \textcolor{blue}{v_j} - x_{kj}\textcolor{blue}{\theta_k} + x_{kj}\theta_k'  & \triangleright\ O(1)\\
	&p_j' \leftarrow \logistic(v_j') & \triangleright\ O(1)\\
	&l_j' \leftarrow y_j\log p_j'+(1-y_j)\log(1-p_j') & \triangleright\ O(1)\\
	&L_k \leftarrow L_k + l_j' & \triangleright\ O(1)\\
	&\textbf{end}\\
	&U_0' \leftarrow \textcolor{blue}{U_0} + \log \pi_0(\textcolor{blue}{\theta_k}) - \log \pi_0(\theta_k') & \triangleright\ O(1)\\
	&U_k\leftarrow L_k + U_0' & \triangleright\ O(1)\\
	&\textbf{Return } U_k
	\]

\begin{figure}[h!]
	\centering
	\begin{subfigure}{0.48\textwidth}
		\centering
		\caption{Compute graph}
		\scalebox{0.8}{
			\begin{tikzpicture}
				\node[state, fill = julia2] (12) {$\theta_2$};
				\node[state] (11) [left = of 12] {$\theta_1$};
				\node[state] (13) [right = of 12] {$\theta_3$};
				
				\node[state, fill = julia2] (21) [below = of 12, xshift = -1cm] {$v_1$};
				\node[state, fill = julia2] (22) [below = of 12, xshift = 1cm] {$v_2$};
				\node[state, fill = julia2] (23) [right = of 22, xshift = 1cm] {$U_0$};
				
				\node[state, fill = julia2] (31) [below = of 21, yshift = 0.5cm] {$l_1$};
				\node[state, fill = julia2] (32) [below = of 22, yshift = 0.5cm] {$l_2$};
				\node[state, fill = julia2] (4) [below = of 31, xshift = 1cm, yshift = 0.5cm] {$L$};
				\node[state, fill = julia2] (5) [below right = of 4, yshift = 0.5cm] {$U$};
				
				\draw (11)--(21);
				\draw (11)--(22);
				\draw (11)--(23);
				\draw[color = julia2] (12)--(21);
				\draw[color = julia2] (12)--(22);
				\draw[color = julia2] (12)--(23);
				\draw (13)--(21);
				\draw (13)--(22);
				\draw (13)--(23);
				
				\draw[color = julia2] (21)--(31);
				\draw[color = julia2] (22)--(32);
				\draw[color = julia2] (31)--(4);
				\draw[color = julia2] (32)--(4);
				\draw[color = julia2] (4)--(5);
				\draw[color = julia2] (23)--(5);
		\end{tikzpicture}}
	\end{subfigure}
	\begin{subfigure}{0.48\textwidth}
		\centering
		\caption{$2^{nd}$ coordinate subgraph}
		\scalebox{0.8}{
			\begin{tikzpicture}
				\node[state] (12) {$\theta_2'$};
				
				\node[state] (21) [below = of 12, xshift = -1cm] {$v_1'$};
				\node[state] (22) [below = of 12, xshift = 1cm] {$v_2'$};
				\node[state] (23) [right = of 22, xshift = 1cm] {$U_0'$};
				
				\node[state] (31) [below = of 21, yshift = 0.5cm] {$l_1'$};
				\node[state] (32) [below = of 22, yshift = 0.5cm] {$l_2'$};
				\node[state] (4) [below = of 31, xshift = 1cm, yshift = 0.5cm] {$L_2$};
				\node[state] (5) [below right = of 4, yshift = 0.5cm] {$U_2$};
				
				\draw (12)--(21);
				\draw (12)--(22);
				\draw (12)--(23);
				
				\draw (21)--(31);
				\draw (22)--(32);
				\draw (31)--(4);
				\draw (32)--(4);
				\draw (4)--(5);
				\draw (23)--(5);
		\end{tikzpicture}}
	\end{subfigure}
	\begin{subfigure}{0.48\textwidth}
		\centering
		\caption{Graphical model}
		\scalebox{0.8}{
			\begin{tikzpicture}
				\node[state] (b2) {$\theta_2,U_0$};
				\node[state] (b1) [left = of b2] {$\theta_1,U_0$};
				\node[state] (b3) [right = of b2] {$\theta_3,U_0$};
				\node[state,fill=lightgray] (y1) [below left = of b2] {$Y_1,l_1$};
				\node[state,fill=lightgray] (y2) [below right = of b2] {$Y_1,l_2$};
				
				\draw (b1)--(y1);
				\draw (b1)--(y2);
				\draw (b2)--(y1);
				\draw (b2)--(y2);
				\draw (b3)--(y1);
				\draw (b3)--(y2);
		\end{tikzpicture}}
	\end{subfigure}
	\caption{
		Different graph notions for logistic regression model in \cref{eg:glm} with 
		three regression parameters and two data points. Each node in the two
		compute graphs corresponds to the values in their pseudocode and each node
		in the graphical model stores the variable name (left symbol) and its conditional
		negative log density (right symbol). The colored nodes and edges on the 
		full graph are used to construct the coordinate subgraph.
		In the figure we compute $U_2$, which costs $O(dn)$ using the
		graphical model and $O(n)$ using the $2^{nd}$ coordinate subgraph.}
	\label{fig:glm_example}
\end{figure}
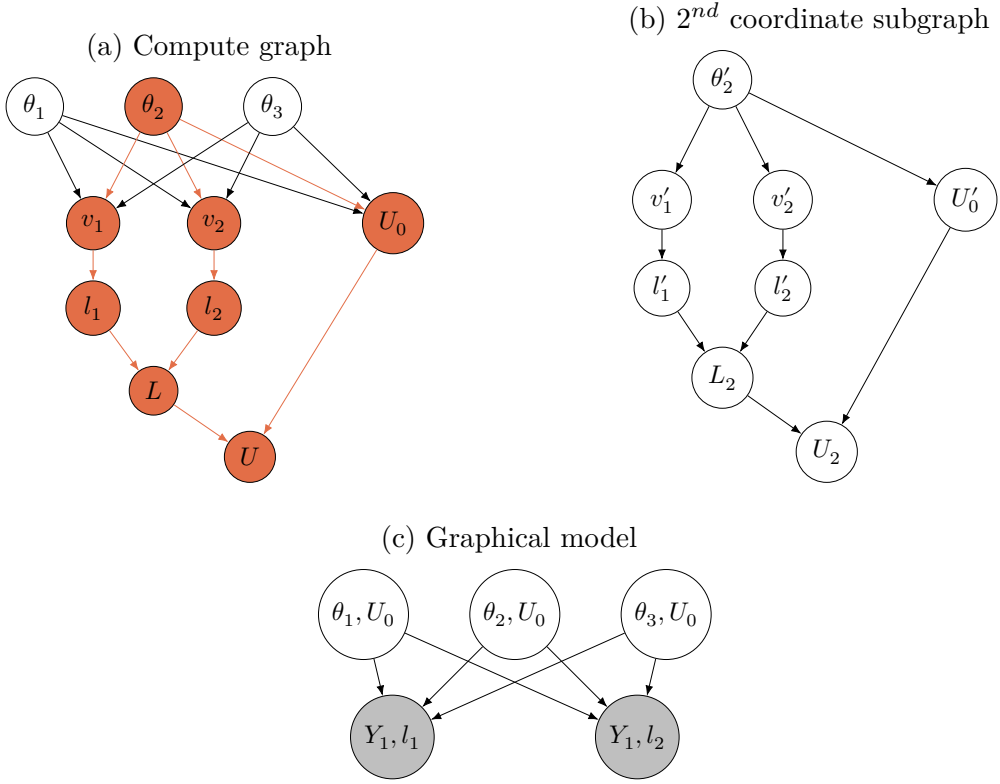

Here, each variable in the pseudocode is a node in the corresponding compute graph or coordinate
subgraph and the stored values in the full graph nodes are highlighted blue when they are used 
in the evaluation of the coordinate subgraph (see \cref{fig:glm_example}). 
From the pseudocode above we can clearly see that 
the full graph costs $c = O(dn)$ and the coordinate subgraph costs $c_k = O(n)$ for all 
$k=1,\ldots,d$. Hence this model is a CF distribution.

On the other hand, from the graphical model in \cref{fig:glm_example}, we obtain the following decomposition  (see, e.g., \citet{jordan_graphical_2004})
\[
\pi(\theta) &\propto p(\theta,y)\\
&\propto \exp\left(- \sum_{j=1}^{n}l_j(\theta) - U_0(\theta_k) \right)\exp\left(- \sum_{j\neq k}U_0(\theta_j) \right).\label{eq:glm_gm}
\]
Note that in \cref{eq:glm_gm}, only the terms inside the first exponent are dependent on $\theta_k$.
This means we can evaluate $U_k(\theta'_k)$ by evaluating $l_j(\theta')$ and $U_0(\theta_k')$.
However, evaluating $l_j(\theta')$ costs $O(dn)$ which is $O(d)$ times more expensive than evaluating
the $k^{th}$ coordinate subgraph. In this case, evaluating $l_j(\theta')$ unnecessarily recomputes
the $v_j,l_j$ and $U_0$ nodes in the full graph.
\eexa

\bexa \label{eg:AR}
Consider the following $AR(d)$ model for time series data with
independent priors $\pi_0$ on the coefficients
\[
X_t &= \sum_{i=1}^{d}\theta_iX_{t-i} + \epsilon_t, & \epsilon_t\sim N(0,\sigma^2)\\
\theta_i &\sim \pi_c, &i=1,\ldots,d
\]
Assuming we have data $x=(x_1,\ldots,x_T)$ and $T>d$, the posterior for such target has the following form
\[
\pi(\theta) &\propto \exp\left(-\frac{1}{2\sigma^2}\sum_{t=d+1}^{T}
\left(x_t - \sum_{i=1}^{d}\theta_ix_{t-i}\right)^2\right)
\prod_{i=1}^{d}\pi_0(\theta_i)\\
&\propto \exp\left[(\theta-\mu)^\top\Sigma(\theta-\mu)\right]\prod_{i=1}^{d}\pi_0(\theta_i)\\
\Rightarrow U(\theta) &= \underbrace{(\theta-\mu)^\top\Sigma(\theta-\mu)}_{L(\theta)} \underbrace{- \sum_{i=1}^{d}\log\pi_0(\theta_i)}_{U_0(\theta)}
\]
where the second line is due to the expression in the exponential being a quadratic form of $\theta$.
Also, since $\mu\in\reals^d$ and $\Sigma = (\Sigma_1\ldots\Sigma_d)\in\reals^{d\times d}$ are fixed quantities and can be
computed at the beginning with a one time $O(T)$ cost, we can ignore this cost in the evaluation.
Again, we look at the pseudocode and cost breakdown for the full graph and coordinate subgraph of 
this target

\begin{multicols}{2}
	\textbf{Full graph evaluation}
	\[
	&v \leftarrow \Sigma(\theta - \mu) & \triangleright\ O(d^2)\\
	&L \leftarrow (\theta - \mu)^\top v & \triangleright\ O(d)\\
	&\text{Set }U_0\leftarrow 0 \\
	&\textbf{For } i=1,\ldots,d & \triangleright\ O(d)\\
	&U_0 \leftarrow U_0 + \log \pi_0(\theta_i) & \triangleright\ O(1)\\
	&\textbf{end}\\
	&U\leftarrow L + U_0 & \triangleright\ O(1)\\
	&\textbf{Return } U
	\]
	
	\columnbreak
	
	\textbf{Coordinate subgraph evaluation}
	\[
	&v_k \leftarrow \textcolor{blue}{v} + (\theta_k'-\theta_k)\Sigma_k & \triangleright\ O(d)\\
	&L_k \leftarrow (\theta - \mu)^\top v_k & \triangleright\ O(d)\\
	&U_0' \leftarrow \textcolor{blue}{U_0} - \log \pi_0(\textcolor{blue}{\theta_k}) + \log \pi_0(\theta_k') & \triangleright\ O(1)\\
	&U_k\leftarrow L_k + U_0' & \triangleright\ O(1)\\
	&\textbf{Return } U_k
	\]
	
\end{multicols}

Here, we see that the full graph costs $c = O(d^2)$ and the coordinate subgraph 
costs $c_k = O(d)$ for all $k=1,\ldots,d$. Hence this model is also a CF distribution.

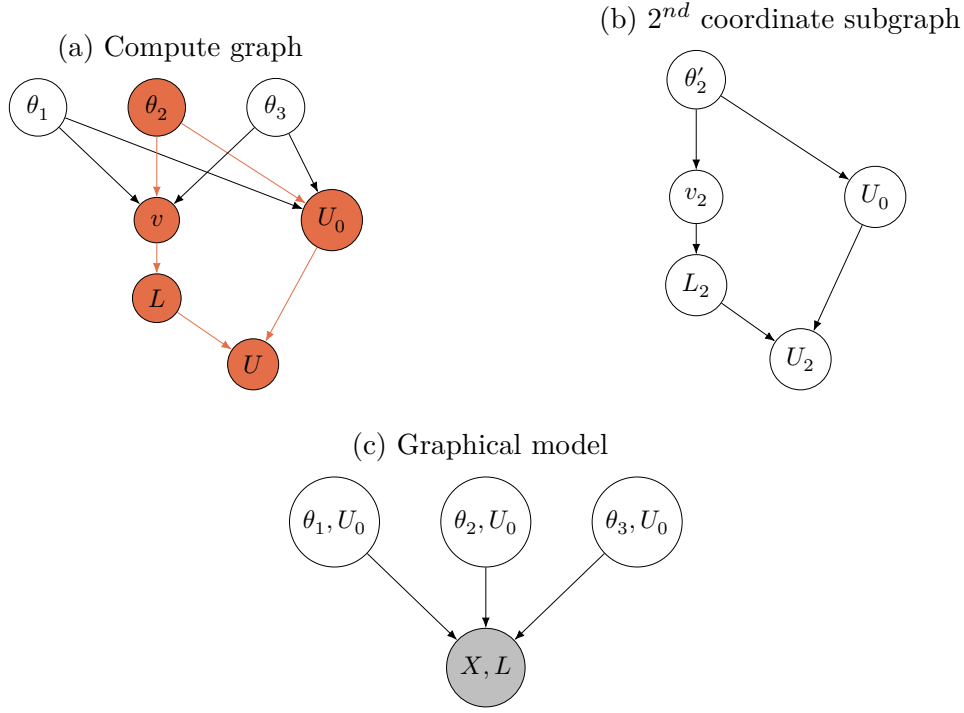
\begin{figure}[h!]
	\centering
	\begin{subfigure}{0.48\textwidth}
		\centering
		\caption{Compute graph}
		\scalebox{0.8}{
			\begin{tikzpicture}
				\node[state, fill = julia2] (12) {$\theta_2$};
				\node[state] (11) [left = of 12] {$\theta_1$};
				\node[state] (13) [right = of 12] {$\theta_3$};
				
				\node[state, fill = julia2] (21) [below = of 12] {$v$};
				\node[state, fill = julia2] (22) [right = of 21, xshift = 1cm] {$U_0$};
				
				\node[state, fill = julia2] (3) [below = of 21, yshift = 0.5cm] {$L$};
				\node[state, fill = julia2] (4) [below right = of 3, yshift = 0.5cm] {$U$};
				
				\draw (11)--(21);
				\draw (11)--(22);
				\draw[color = julia2] (12)--(21);
				\draw[color = julia2] (12)--(22);
				\draw (13)--(21);
				\draw (13)--(22);
				
				\draw[color = julia2] (21)--(3);
				\draw[color = julia2] (3)--(4);
				\draw[color = julia2] (22)--(4);
		\end{tikzpicture}}
	\end{subfigure}
	\begin{subfigure}{0.48\textwidth}
		\centering
		\caption{$2^{nd}$ coordinate subgraph}
		\scalebox{0.8}{
			\begin{tikzpicture}
				\node[state] (12) {$\theta_2'$};
				
				\node[state] (21) [below = of 12] {$v_2$};
				\node[state] (22) [right = of 21, xshift = 1cm] {$U_0$};
				
				\node[state] (3) [below = of 21, yshift = 0.5cm] {$L_2$};
				\node[state] (4) [below right = of 3, yshift = 0.5cm] {$U_2$};
				
				\draw (12)--(21);
				\draw (12)--(22);
				
				\draw (21)--(3);
				\draw (3)--(4);
				\draw (22)--(4);
		\end{tikzpicture}}
	\end{subfigure}
	\begin{subfigure}{0.48\textwidth}
		\centering
		\caption{Graphical model}
		\scalebox{0.8}{
			\begin{tikzpicture}
				\node[state] (b2) {$\theta_2,U_0$};
				\node[state] (b1) [left = of b2] {$\theta_1,U_0$};
				\node[state] (b3) [right = of b2] {$\theta_3,U_0$};
				\node[state,fill=lightgray] (y1) [below = of b2] {$X,L$};
				
				\draw (b1)--(y1);
				\draw (b2)--(y1);
				\draw (b3)--(y1);
		\end{tikzpicture}}
	\end{subfigure}
	\caption{
		Different graph notions for autoregression model in \cref{eg:AR} with 
		three autoregression parameters. Each node in the two
		compute graphs corresponds to the values in their pseudocode and each node
		in the graphical model stores the variable name (left symbol) and its conditional
		negative log density (right symbol). The colored nodes and edges on the 
		full graph are used to construct the coordinate subgraph.
		In the figure we compute $U_2$, which costs $O(dn)$ using the
		graphical model and $O(n)$ using the $2^{nd}$ coordinate subgraph.}
	\label{fig:AR_example}
\end{figure}

Using the same reasoning as the previous example,
\[
\pi(\theta) &\propto \exp\left(- L(\theta) - U_0(\theta_k) \right)\exp\left(- \sum_{j\neq k}U_0(\theta_j) \right).\label{eq:AR_gm}
\]
In \cref{eq:AR_gm}, only the terms inside the first exponent are dependent on $\theta_k$.
This means we can evaluate $U_k(\theta'_k)$ by evaluating $L(\theta')$ and $U_0(\theta_k')$.
Again, evaluating $L(\theta')$ costs $O(d^2)$ which is $O(d)$ times more expensive than evaluating
the $k^{th}$ coordinate subgraph due to the recomputing of the $v$ node in the full graph.
\eexa 
\section{Details of experiments}\label{app:exp}
\if1\anon
{
In this section, we provide some additional experiments and details on the 
experiments found in the main text of the paper. All experiments were conducted on the ARC Sockeye computer cluster at the University of British Columbia.
} \fi

\if0\anon
{
In this section, we provide some additional experiments and details on the 
experiments found in the main text of the paper. All experiments were conducted on the [anonymized name] computer cluster at the [anonymized research institute].
} \fi

The license information for used assets is as follows:
\begin{itemize}
\item \textbf{Datasets}: We used the binary classification datasets from UCI Machine Learning Repository,
licensed under the Creative Commons Attribution 4.0 International (CC BY 4.0) license.

\item \textbf{Julia package Pigeons:} GNU Affero General Public License Version 3 (AGPL-3.0).

\item \textbf{Julia package AdvancedHMC:} MIT License
\end{itemize}

\paragraph{Data details} We summarize the dimensions of the real datasets used
in \cref{sec:realdata_experiment} in \cref{tab:datasets}. 

\begin{table}[t]
  \centering
  \caption{Datasets considered in our experiments.}
  \label{tab:datasets}
  \begin{tabular}{llll}
  \hline
  \textbf{Dataset} & \textbf{Sample size} & \textbf{Features} & \textbf{Source}\\
  \hline
  prostate & 102 & 50 (reduced from 5966) & \citep{biron2024automala}(full dataset \citep{singh2002gene})\\
  ionosphere & 351 & 34 & \citep{sigillito1989ionosphere}\\
  sonar & 208 & 60 & \citep{sejnowski1988connectionist}\\
  \hline
  \end{tabular}
\end{table}

\paragraph{Data preprocessing} 

For the prostate cancer dataset,
since the data dimension is large (5966 covariates),
we opt to reduce the number of covariates to 50 to ensure convergence of NRPT in a reasonable amount of time. To do this, we use the dataset processed by \citet{biron2024automala} by first selecting the 25 most important covariates using LASSO, then randomly adding the remaining 25 from the
covariates not selected by LASSO. 

For all datasets, we standardize the design matrix obtained after any dimension reduction
and subsampling procedures are applied. 

\subsection{Additional plots for prostate cancer data experiment}

\begin{figure}[t]
  \centering
  \begin{subfigure}{0.325\textwidth}
    \centering
    \includegraphics[width=\textwidth]{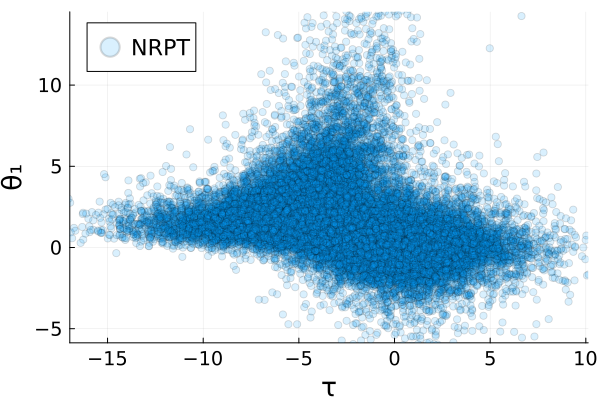}
  \end{subfigure}
  \begin{subfigure}{0.325\textwidth}
    \centering
    \includegraphics[width=\textwidth]{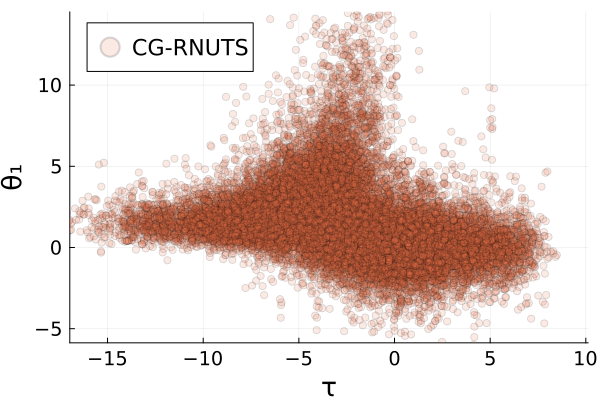}
  \end{subfigure}
  \begin{subfigure}{0.325\textwidth}
    \centering
    \includegraphics[width=\textwidth]{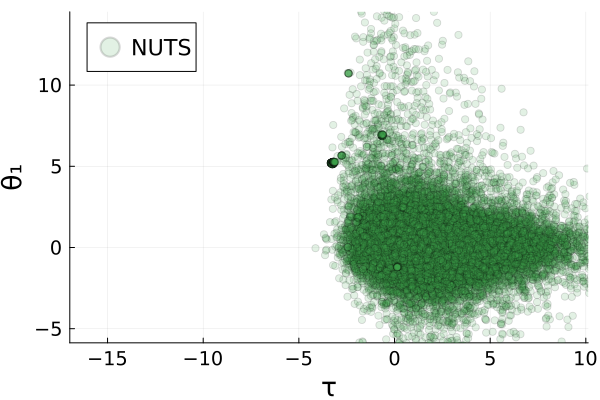}
  \end{subfigure}

  \begin{subfigure}{0.325\textwidth}
    \centering
    \includegraphics[width=\textwidth]{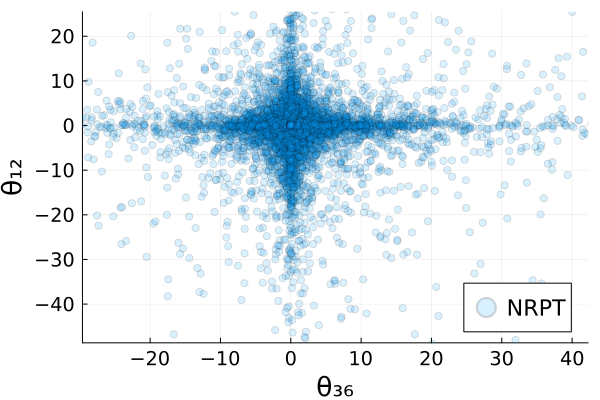}
  \end{subfigure}
  \begin{subfigure}{0.325\textwidth}
    \centering
    \includegraphics[width=\textwidth]{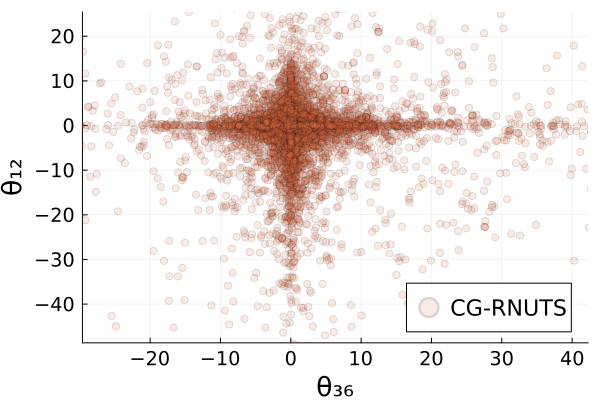}
  \end{subfigure}
  \begin{subfigure}{0.325\textwidth}
    \centering
    \includegraphics[width=\textwidth]{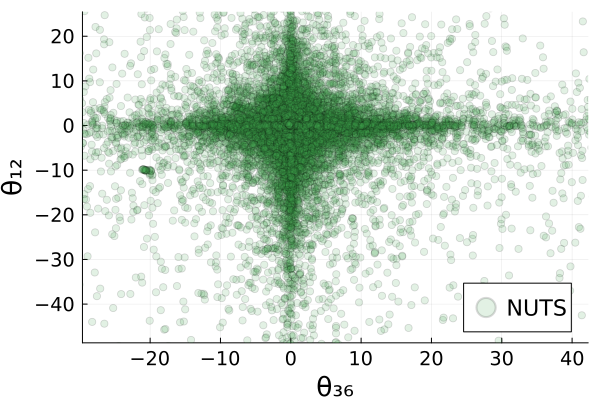}
  \end{subfigure}
  
  \begin{subfigure}{0.325\textwidth}
    \centering
    \includegraphics[width=\textwidth]{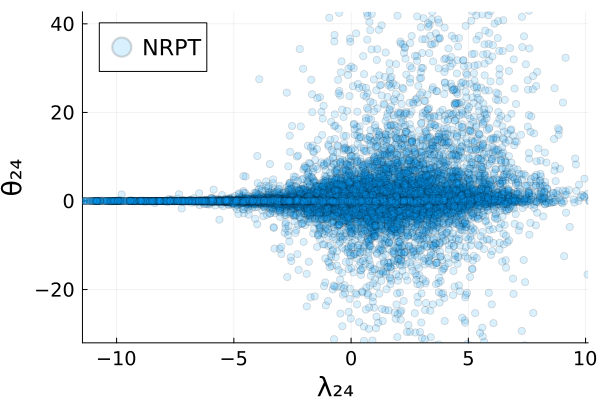}
  \end{subfigure}
  \begin{subfigure}{0.325\textwidth}
    \centering
    \includegraphics[width=\textwidth]{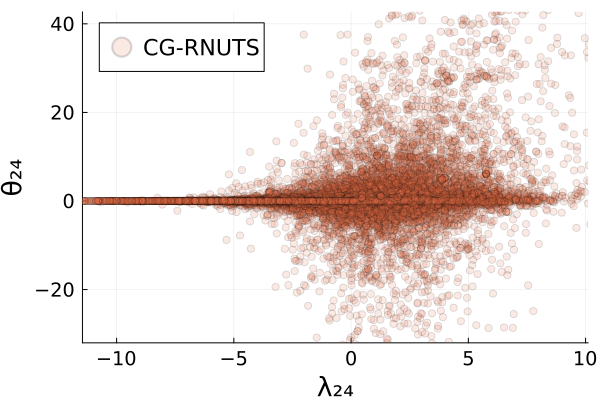}
  \end{subfigure}
  \begin{subfigure}{0.325\textwidth}
    \centering
    \includegraphics[width=\textwidth]{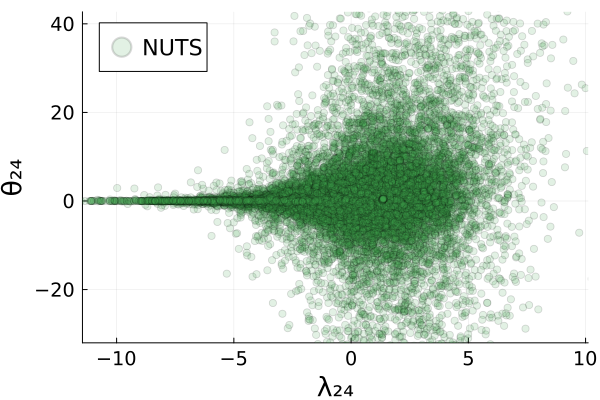}  
  \end{subfigure}

  \begin{subfigure}{0.325\textwidth}
    \centering
    \includegraphics[width=\textwidth]{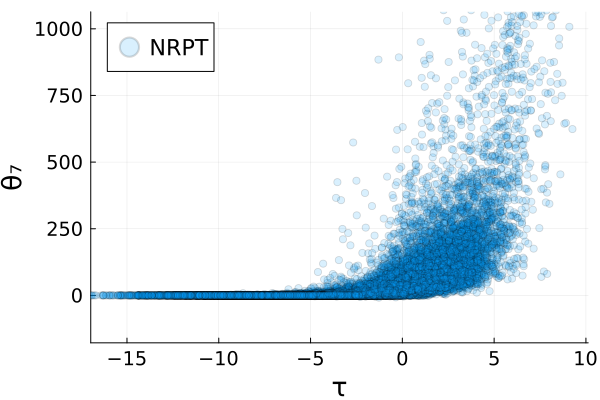}
  \end{subfigure}
  \begin{subfigure}{0.325\textwidth}
    \centering
    \includegraphics[width=\textwidth]{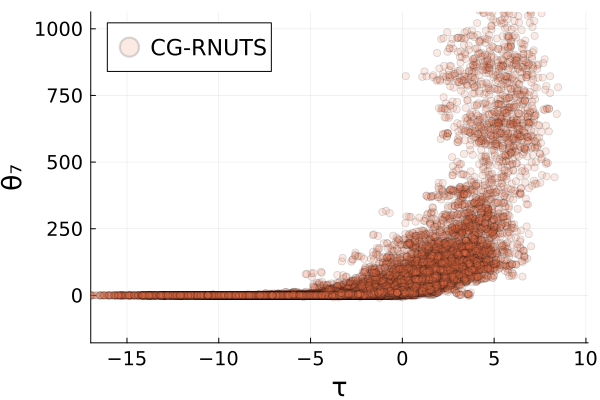}
  \end{subfigure}
  \begin{subfigure}{0.325\textwidth}
    \centering
    \includegraphics[width=\textwidth]{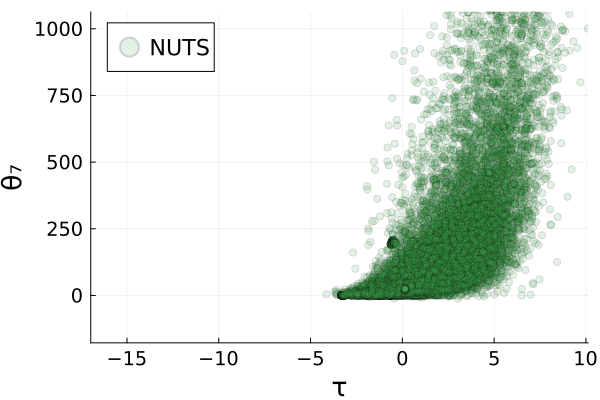}
  \end{subfigure}
  
  \begin{subfigure}{0.325\textwidth}
    \centering
    \includegraphics[width=\textwidth]{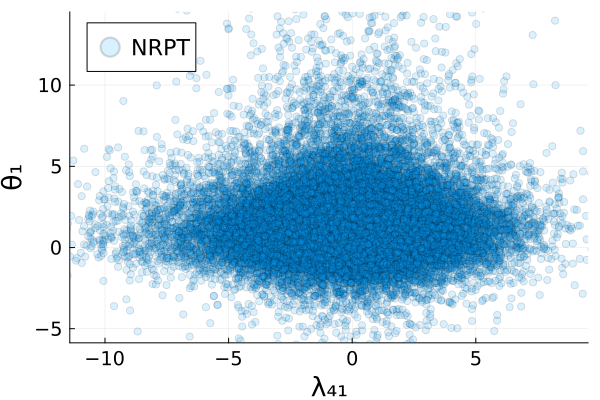}
  \end{subfigure}
  \begin{subfigure}{0.325\textwidth}
    \centering
    \includegraphics[width=\textwidth]{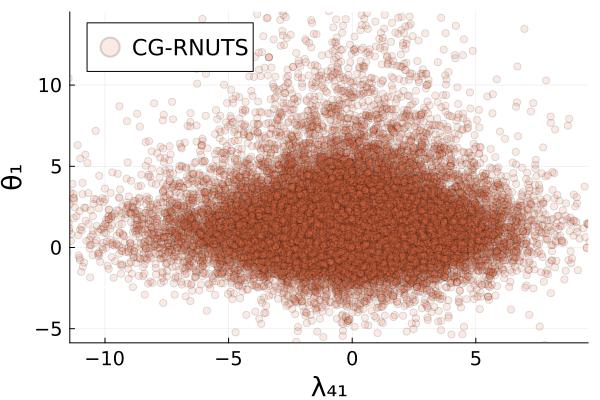}
  \end{subfigure}
  \begin{subfigure}{0.325\textwidth}
    \centering
    \includegraphics[width=\textwidth]{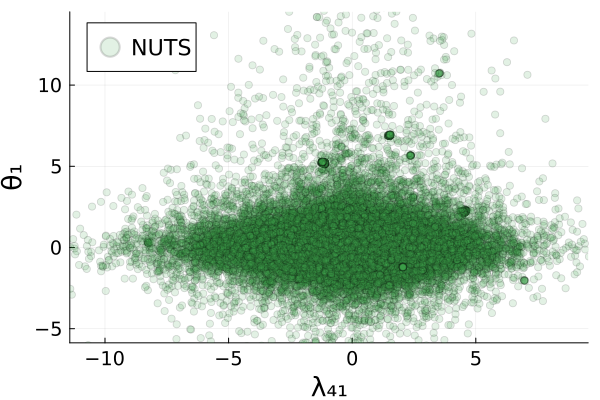}
  \end{subfigure}

  \caption{Bivariate marginals obtained by CG-RNUTS and NUTS with similar time budgets compared to
  the reference posterior obtained by NRPT for
  the horseshoe logistic regression model parameters on prostate cancer data with a subsample of size 10. 
  \textbf{(Left column)} NRPT (used as reference posterior), \textbf{(middle column)} CG-RNUTS, 
  \textbf{(right column)} NUTS.}
  \label{fig:horseshoe_pairplot} 
\end{figure}

In this subsection, we show the bivariate marginals of different pairs
of model parameters obtained by NRPT, CG-RNUTS and NUTS from the prostate cancer data 
experiment in \cref{sec:realdata_experiment}. From \cref{fig:horseshoe_pairplot}, we can 
see the variety of geometry structures that the horseshoe logistic regression model can
exhibit and how NUTS can fail at exploring many of them.

\begin{figure}[t]
  \centering
  \begin{subfigure}{0.325\textwidth}
    \centering
    \includegraphics[width=\textwidth]{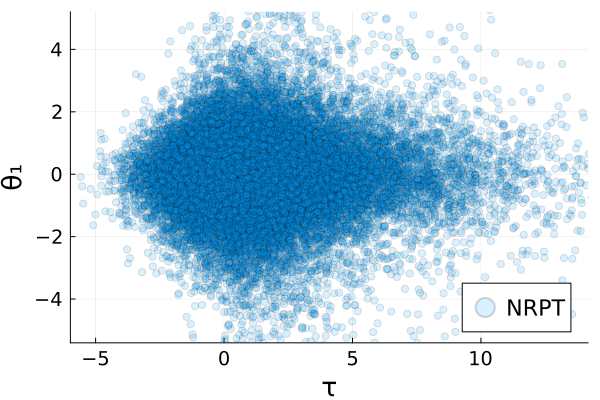}
  \end{subfigure}
  \begin{subfigure}{0.325\textwidth}
    \centering
    \includegraphics[width=\textwidth]{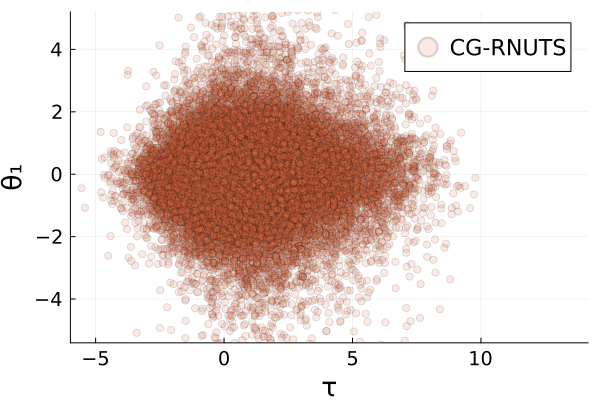}
  \end{subfigure}
  \begin{subfigure}{0.325\textwidth}
    \centering
    \includegraphics[width=\textwidth]{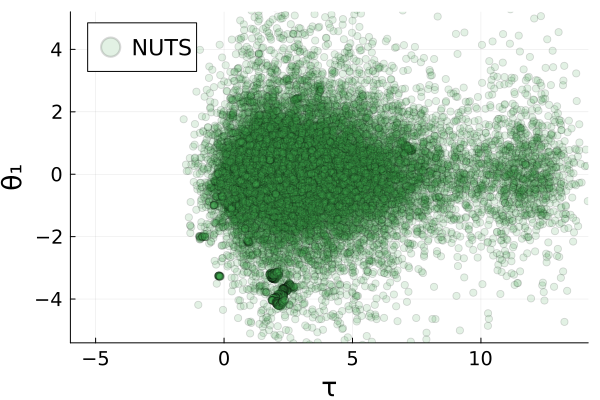}
  \end{subfigure}

  \begin{subfigure}{0.325\textwidth}
    \centering
    \includegraphics[width=\textwidth]{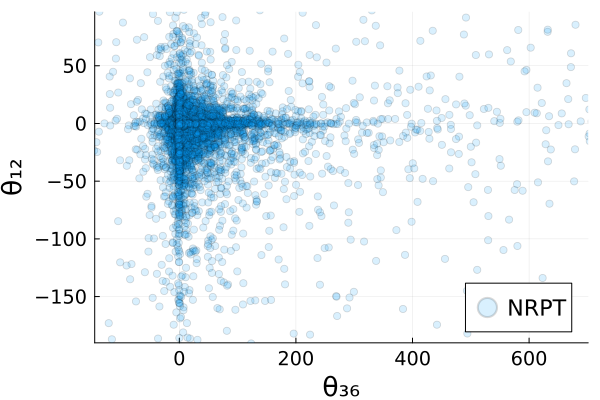}
  \end{subfigure}
  \begin{subfigure}{0.325\textwidth}
    \centering
    \includegraphics[width=\textwidth]{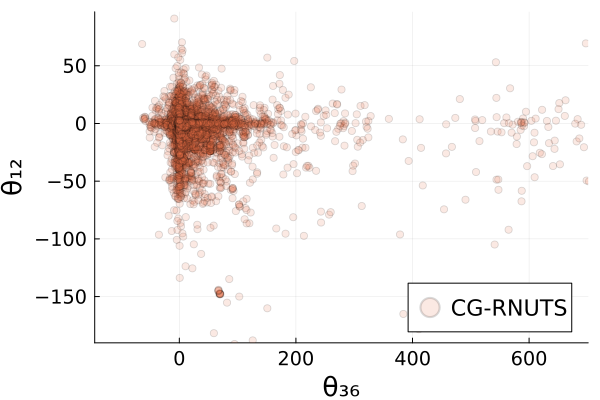}
  \end{subfigure}
  \begin{subfigure}{0.325\textwidth}
    \centering
    \includegraphics[width=\textwidth]{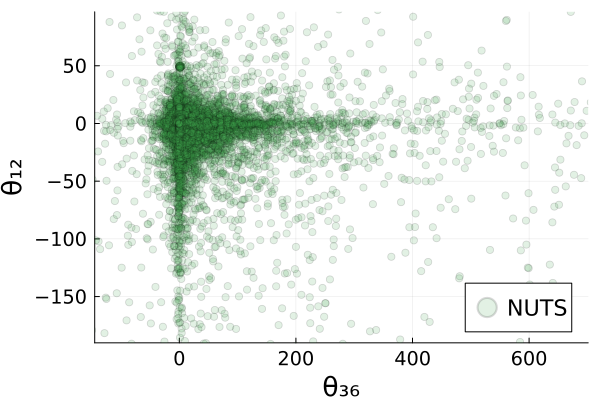}
  \end{subfigure}
  
  \begin{subfigure}{0.325\textwidth}
    \centering
    \includegraphics[width=\textwidth]{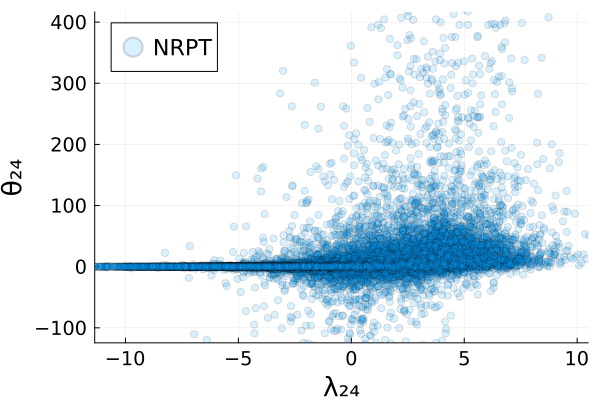}
  \end{subfigure}
  \begin{subfigure}{0.325\textwidth}
    \centering
    \includegraphics[width=\textwidth]{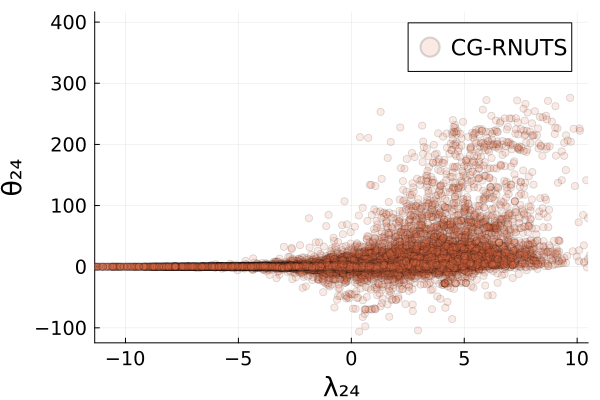}
  \end{subfigure}
  \begin{subfigure}{0.325\textwidth}
    \centering
    \includegraphics[width=\textwidth]{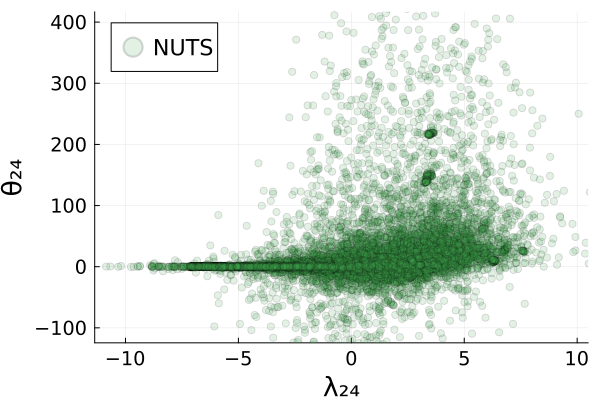}  
  \end{subfigure}

  \begin{subfigure}{0.325\textwidth}
    \centering
    \includegraphics[width=\textwidth]{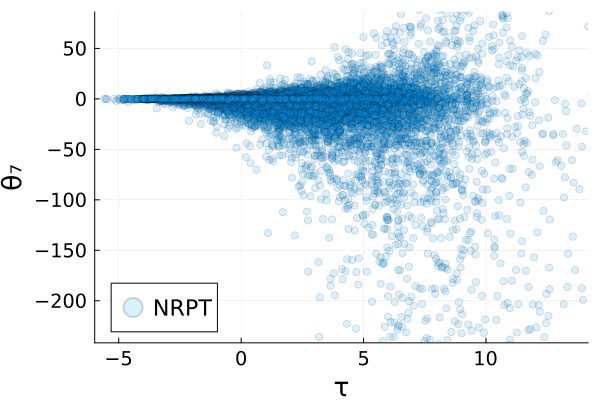}
  \end{subfigure}
  \begin{subfigure}{0.325\textwidth}
    \centering
    \includegraphics[width=\textwidth]{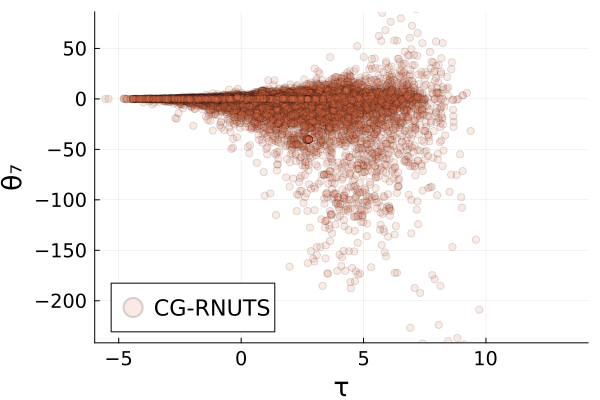}
  \end{subfigure}
  \begin{subfigure}{0.325\textwidth}
    \centering
    \includegraphics[width=\textwidth]{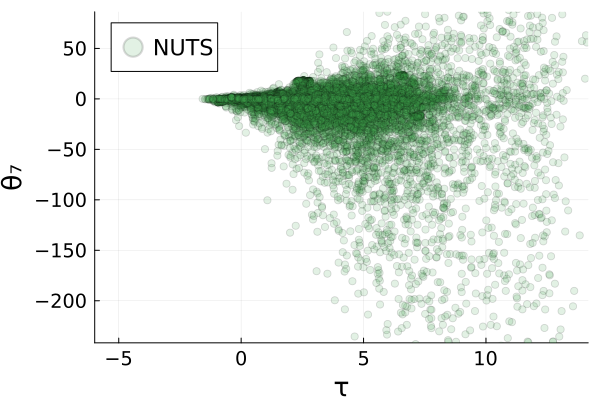}
  \end{subfigure}
  
  \begin{subfigure}{0.325\textwidth}
    \centering
    \includegraphics[width=\textwidth]{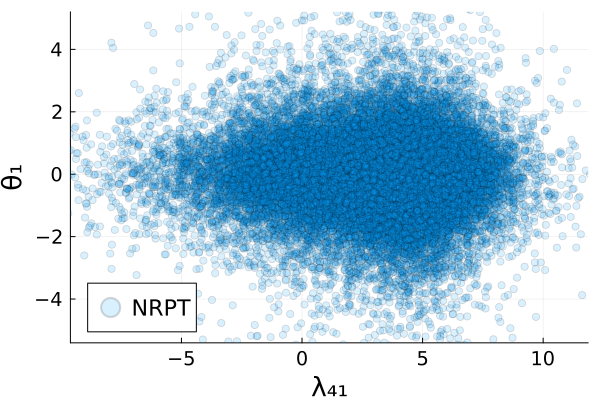}
  \end{subfigure}
  \begin{subfigure}{0.325\textwidth}
    \centering
    \includegraphics[width=\textwidth]{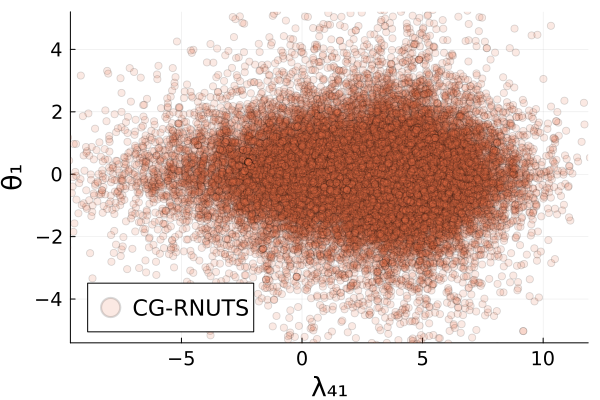}
  \end{subfigure}
  \begin{subfigure}{0.325\textwidth}
    \centering
    \includegraphics[width=\textwidth]{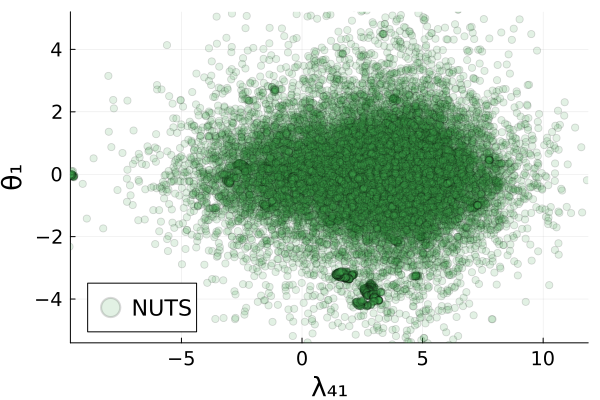}
  \end{subfigure}

  \caption{Bivariate marginals obtained by CG-RNUTS and NUTS with similar time budgets compared to
  the reference posterior obtained by NRPT for
  the horseshoe logistic regression model parameters on prostate cancer data with no subsampling. 
  \textbf{(Left column)} NRPT (used as reference posterior), \textbf{(middle column)} CG-RNUTS, 
  \textbf{(right column)} NUTS.}
  \label{fig:horseshoe_pairplot_full}
\end{figure}

Finally, we show in \cref{fig:horseshoe_pairplot_full} the bivariate marginals of the same parameters for the horseshoe logistic
regression model for the prostate cancer data with no subsampling (corresponding to the prostate
full target in the panel of datasets experiment). Here, we see that the posterior exhibits 
less diverse geometries compared to the subsampled posterior. This justifies our decision 
to perform subsampling on the dataset.

\subsection{Additional plots for panel of datasets experiment}

\begin{figure}[t]
  \centering
  \begin{subfigure}{0.80\textwidth}
    \centering
    \includegraphics[width=0.95\textwidth]{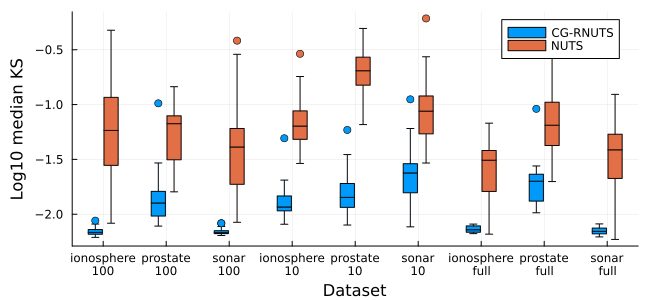}
  \end{subfigure}
  \begin{subfigure}{0.80\textwidth}
    \centering
    \includegraphics[width=0.95\textwidth]{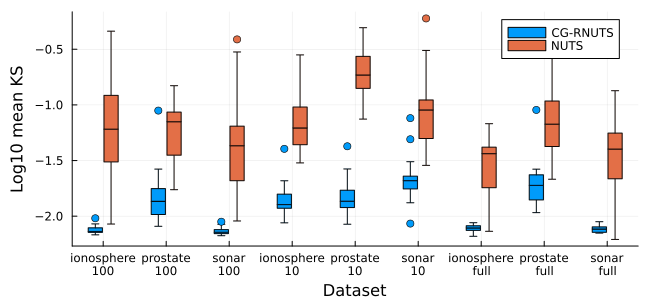}
  \end{subfigure}
  \begin{subfigure}{0.80\textwidth}
    \centering
    \includegraphics[width=0.95\textwidth]{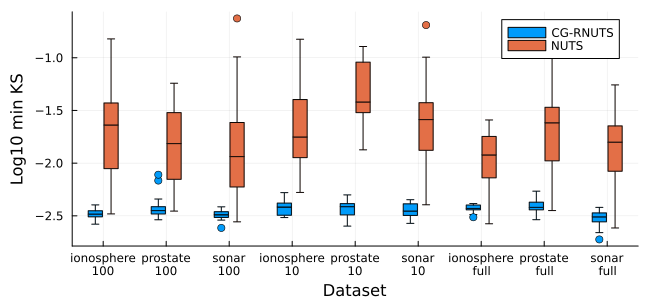}
  \end{subfigure}
  \caption{Box plot of median (top), mean (middle) and minimum (bottom) marginal KS statistics 
  of CG-RNUTS and NUTS compared to the NRPT benchmark (lower is better) for 
  horseshoe logistic regression targets with various datasets and data sizes.}
  \label{fig:panel_stats}
\end{figure}

In this subsection, we show the median, mean and minimum marginal KS statistics
of CG-RNUTS and NUTS compared to the NRPT benchmark for \cref{sec:panel_dataset}'s
panel of datasets experiment in \cref{fig:panel_stats}. As expected, CG-RNUTS outperforms 
NUTS in all metrics and settings considered.

\begin{figure}[t]
  \centering
  \begin{subfigure}{0.80\textwidth}
    \centering
    \includegraphics[width=0.95\textwidth]{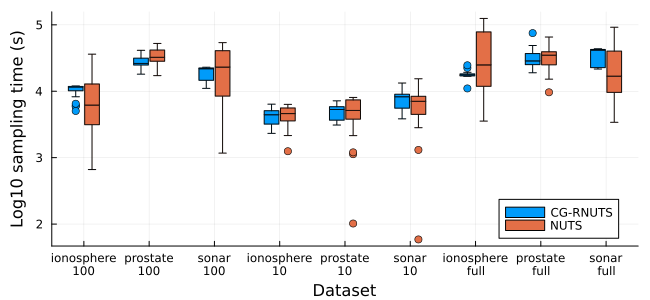}
  \end{subfigure}
  \begin{subfigure}{0.80\textwidth}
    \centering
    \includegraphics[width=0.95\textwidth]{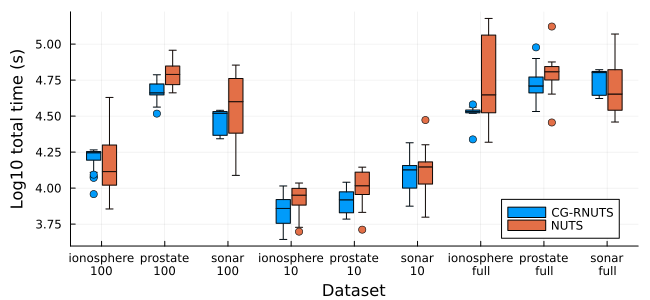}
  \end{subfigure}
  \caption{Box plot of sampling time (left) and total time (right) for CG-RNUTS and NUTS 
  for horseshoe logistic regression targets with various datasets and data sizes.}
  \label{fig:panel_time}
\end{figure}

We also show the sampling and total time for each algorithm in \cref{fig:panel_time}
to confirm CG-RNUTS and NUTS were given similar time budgets.

\section{Modes of Automatic Differentiation} \label{app:AD_modes}

In this section, we review different modes of 
Automatic Differentiation (AD) and how they are used to 
compute different differential quantities in practice.
Consider a function $f:\reals^d\to\reals^m$ whose derivatives
we want to compute. Denote $\nabla f\in \reals^d\times \reals^m$
the Jacobian matrix of $f$ and $c=c(d)$ the evaluation cost of $f$. 
Additionally, denote $e_i$ the $i^{th}$ coordinate vector, i.e. a vector with the $i^{th}$
element equal to 1 and 0 elsewhere. 

\paragraph{Forward mode AD} This AD mode evaluates the Jacobian
vector product of $f$, i.e.,
\[
\texttt{\_forwardAD}(f,x,v) = \nabla f(x) v,
\]
where $v$ is the vector indicating the derivative direction.
To evaluate \verb|_forwardAD|, we simply evaluate the compute
graph of $f$ but with input $x + \varepsilon v$ where $\varepsilon$ is 
the abstract number with property $\varepsilon^2 = 0$. Note
that evaluation for this type of input is similar to evaluating
functions of complex numbers but with the arithmetic rule of 
$\varepsilon$. At the end of the evaluation, the coefficients of 
$\varepsilon$ in the sink nodes are the outputs of \verb|forwardAD|.
Each application of \verb|_forwardAD| costs $O(c)$ to compute.

When $m=1$, we can apply \verb|_forwardAD|$(f,x,e_i)$ for $i=1,\ldots,d$
to get the gradient of $f$. Then, we get the definition for the 
\verb|forwardAD| function in \cref{sec:hess_trick}
\[
\forwardAD(f,x) := (\texttt{\_forwardAD}(f,x,e_1),\ldots,\texttt{\_forwardAD}(f,x,e_d)).
\]
Since \verb|forwardAD| is $d$ applications of \verb|_forwardAD|, it costs $O(dc)$
per application.

\paragraph{Reverse mode AD} This AD mode evaluates the vector Jacobian
product of $f$, i.e.,
\[
\texttt{\_reverseAD}(f,x,v) = v^\top\nabla f(x),
\]
where $v$ is the vector indicating the output linear combination. 
In other words, reverse mode AD computes the gradient of $v^\top f$.
To implement this mode of AD, we first do a forward pass of the 
compute graph at $x$ then do a reverse pass to get the result. 
Here a reverse pass is done by travelling up the compute graph
starting from the sink nodes with values 1 and use the derivative
chain rule to compute the values in the parent nodes. At the end
of the reverse pass, the value we get in each input node is
the partial derivative value with respect to that input at $x$.
Since the forward and reverse pass both costs $O(c)$, the complexity
of \verb|_reverseAD| is $O(c)$. The \verb|reverseAD| function in
the main text is a special case where $m=1,v=1$ and therefore also
costs $O(c)$
\[
\reverseAD(f,x) := \texttt{\_reverseAD}(f,x,1).
\]

\paragraph{Jacobian AD} When $d=m$, most AD libraries compute the Jacobian
of $f$ by \verb|_forwardAD|$(f,x,e_i)$ for $i=1,\ldots,d$, i.e.,
\[
\JacobianAD(f,x) := (\texttt{\_forwardAD}(f,x,e_1),\ldots,\texttt{\_forwardAD}(f,x,e_d)).
\]
Similar to \verb|forwardAD|, the \verb|JacobianAD| function costs $O(dc)$ to evaluate.
To simplify things, we use the following definition of \verb|JacobianAD| for functions
whose output is a diagonal matrix:
\[
\JacobianAD(\diag(f_1(\cdot),\ldots,f_d(\cdot)),x) := \JacobianAD((f_1(\cdot),\ldots,f_d(\cdot)),x),
\]
where $f_i(\cdot)$ are $\reals\to\reals$ functions.

\paragraph{Hessian AD} When $m=1$, computing the Hessian of $f$ is a special case of Jacobian
computation where the \verb|JacobianAD| is applied to the reverse mode gradient of $f$, i.e.,
\[
\HessianAD(f,x) := \JacobianAD(\reverseAD(f,\cdot),x).
\]
In this case, computing $\reverseAD(f,\cdot)$ costs $O(c)$ so \verb|HessianAD|
costs $O(dc)$.
\section{Implementation considerations for RHMC}\label{app:RHMC_considerations}

In this section, we discuss other implementation considerations for RHMC besides
the evaluation of the metric $M(\theta)$.

\subsection{Termination criterion for the fixed-point routine} \label{app:fp_routine}

Termination criteria for fixed-point routines is seldom discussed in the
RHMC literature with many works opting for a fixed number of fixed-point
iterations \citep{girolami2011riemann,xu2024practical}. \citet{brofos2021evaluating}
proposed a routine to stop after the maximum norm error between iterations
falls below a pre-specified threshold, $\delta$, and in a later paper \citep{brofos2021numerical} showed empirically
that, for $\delta < 0.01$, many convergence and mixing metrics show diminishing returns
per computational effort. Therefore, we use Algorithm 1 from \citet{brofos2021evaluating} with 
tolerance $\delta = 0.01$ and minimum number of fixed-point iterations, $K_{\min}$, 
set to 6 as a buffer. We set the maximum
number of fixed-point iterations, $K_{\max}$, to 50.

\begin{figure}[t]
  \begin{subfigure}{0.48\textwidth}
    \centering
    \includegraphics[width=0.95\textwidth]{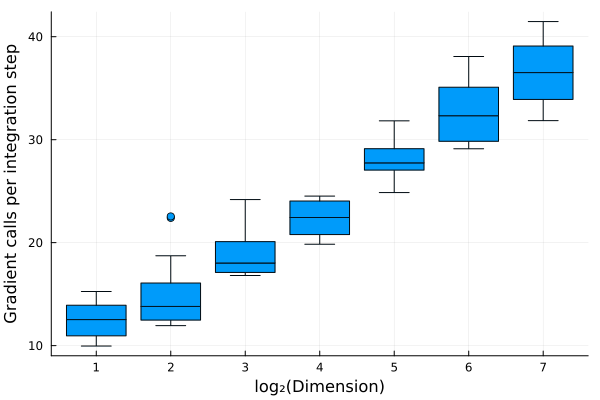}
  \end{subfigure}
  \begin{subfigure}{0.48\textwidth}
    \centering
    \includegraphics[width=0.95\textwidth]{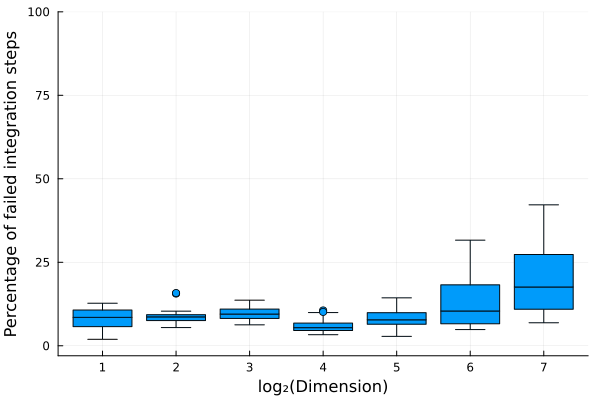}
  \end{subfigure}
  \caption{Fixed-point iteration scaling for funnel targets with increasing dimensions. 
  \textbf{(Left)} Gradient calls (fixed-point iterations) per integration step, 
  \textbf{(right)} Percentage of integration steps that reached $K_{\max}$.}
  \label{fig:fp_scale_funnel}
\end{figure}

To see why the number of fixed-point iterations must be set dynamically, we run CG-RNUTS
on a sequence of funnel targets from the funnel experiment but with increasing dimension 
$d\in \{2^1,\ldots,2^7\}$ and record the average number of fixed-point iterations
(represented by the number of gradient calls) per integration step 
as well as the percentage of failed integration steps (i.e., integration
steps reaching $K_{\max}$ fixed-point iterations). As seen in \cref{fig:fp_scale_funnel}, the
number of fixed-point iterations and integrator failures can be well beyond the recommended
6 iterations from past literature \citep{girolami2011riemann,xu2024practical}.

\begin{figure}[t]
  \begin{subfigure}{0.48\textwidth}
    \centering
    \includegraphics[width=0.95\textwidth]{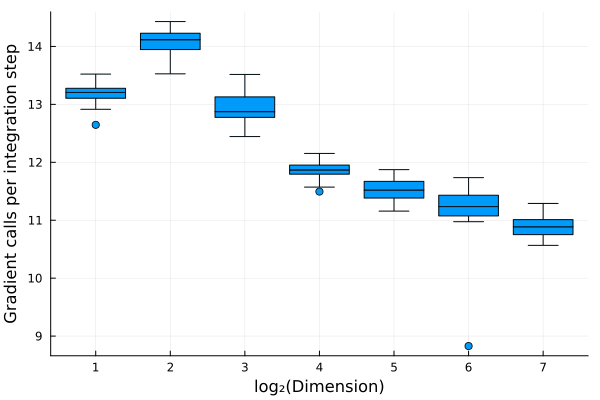}
  \end{subfigure}
  \begin{subfigure}{0.48\textwidth}
    \centering
    \includegraphics[width=0.95\textwidth]{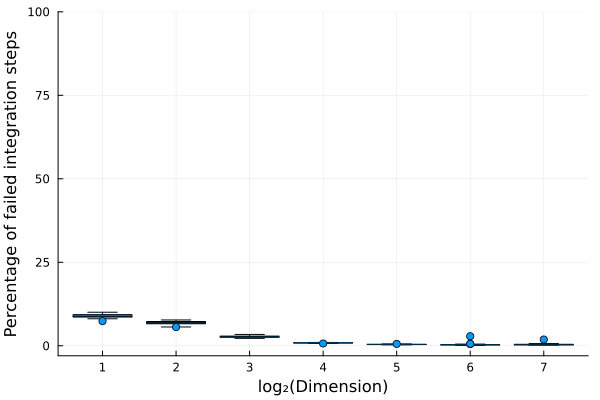}
  \end{subfigure}
  \caption{Fixed-point iteration scaling for synthetic horseshoe logistic regression targets 
  with increasing dimensions. 
  \textbf{(Left)} Gradient calls (fixed-point iterations) per integration step, 
  \textbf{(right)} Percentage of integration steps that reached $K_{\max}$.}
  \label{fig:fp_scale_glm}
\end{figure}

Note however that fixed-point iterations do not always scale with dimension but mostly
depend on the geometry of the target. To see this, we look at the same experiment but
with a sequence of synthetic horseshoe logistic regression targets with increasing dimension 
$d\in \{2^2,\ldots,2^8\}$. In \cref{fig:fp_scale_glm}, we see that the number of fixed-point iterations per integration step does not scale but stabilizes as the dimension of
the target increases and there are very few integrator failures. All that is to say,
we need both a way to dynamically select the number of fixed-point steps for each integration
step and a reasonable stopping criterion to prevent infinite looping of fixed-point iterations.

The last concern with the fixed-point routine is that the numerical errors caused by it 
can potentially violate the involutive and volume preserving properties of the
implicit midpoint integrator. Fortunately, these violations have been shown to be negligible in the same line of previous work  \citep{brofos2021evaluating,brofos2021numerical}.

\subsection{Limitations of the softabs regularizer} \label{app:RHMC_neg_hess}
\begin{figure}[t]
  \centering
  \includegraphics[width=0.7\textwidth]{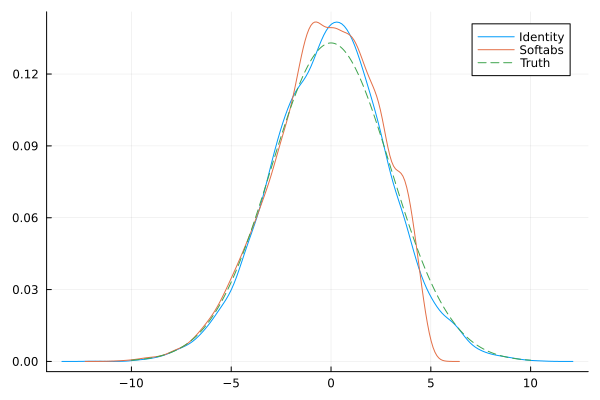}
  \caption{Density of samples from RNUTS chains using different metric 
  regularizations on a 100-dimensional funnel target.}
  \label{fig:softabs_dens}
\end{figure}

\begin{figure}[t]
  \begin{subfigure}{0.48\textwidth}
    \centering
    \includegraphics[width=0.95\textwidth]{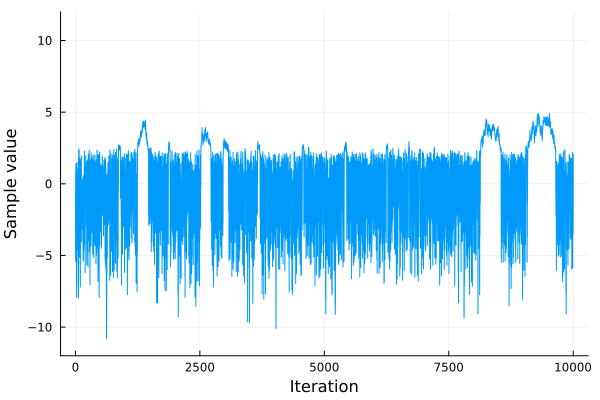}
  \end{subfigure}
  \begin{subfigure}{0.48\textwidth}
    \centering
    \includegraphics[width=0.95\textwidth]{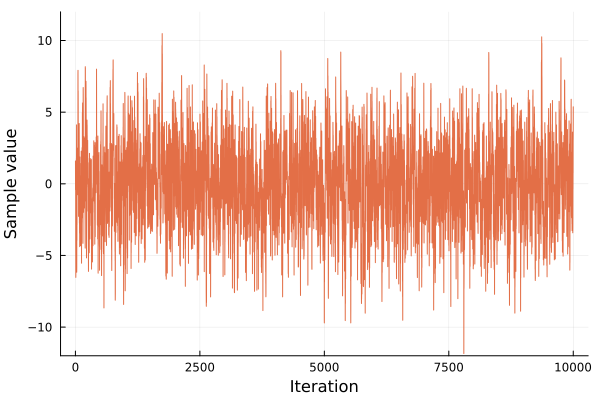}
  \end{subfigure}
  \caption{Trace plots of RNUTS chains using different mass matrix 
  regularizations on a 100-dimensional funnel target. 
  \textbf{(Left)} softabs regularization, \textbf{(right)} no regularization.}
  \label{fig:softabs_trace}
\end{figure}

As mentioned in \cref{sec:riemannian}, the softabs regularizer can prevent exploration
of low curvature regions since it places a lower bound on the metric values.
To see this, consider the same 100-dimensional funnel target in \cref{sec:experiment}. 
Now we run two RNUTS chains with different mass matrices: 
the first one uses $\text{softabs}_{100}$,  
the second one uses the identity regularizer. Next, run each chain for 20000 iterations with
half of those used to adapt the step size. Based on the density plot of 
the two chains in \cref{fig:softabs_dens}, we notice that the right tail
of the softabs chain does not fit the true distribution well. Based on the 
trace plots in \cref{fig:softabs_trace}, we can see that there seems to be a threshold
keeping samples from crossing into regions of higher $x_1$ values for the
softabs chain while the other chain has no problem mixing. Note that in the case of a diagonal Hessian metric,  the funnel target does not require the use of softabs, hence in the experiments in the main text we only use softabs when it is necessary (namely, for the horseshoe logistic regression target).

\section{Suboptimal implementation of the Riemannian Hamiltonian gradient}\label{app:suboptimal}

Algorithm~\ref{algo:suboptimal} provides the detailed runtime cost analysis for the suboptimal method described in Section~\ref{sec:comparison}.

\begin{algorithm}
	\caption{Suboptimal implementation of the Riemannian Hamiltonian gradient}\label{algo:suboptimal}
	\begin{algorithmic}
		\Require \\
		diagonal mass matrix evaluation function \M\ with evaluation cost $c_M$\\
		phase state $z = (\theta, p)$
		\Ensure gradient of Hamiltonian $\nabla H_R(z)$
		\Function{\NaiveGradH}{$\M, z$}
		\State $M \gets \M(\theta)$ \Comment{$O(c_M)$}
		\State $(G_p)_i \gets p_i / M_{i,i},\text{ for all }i \in \{1, 2, \dots, d\}$ \Comment{$O(d)$}
		\State $J \gets \JacobianAD((\M_{1,1}(\cdot), \M_{2,2}(\cdot), \dots, \M_{d,d}(\cdot)),\theta)$  \Comment{$O(dc_M)$}
		\State $G \gets \reverseAD(U, \theta)$ \Comment{$O(c)$}
		\For{$i=1,\ldots,d$}\Comment{$O(d^2)$}
		\State $G_{\theta_i} \gets G_i + \frac{1}{2}\sum_{j=1}^{d}
		J_{j,i}\left( \frac{1}{M_{j,j}} + (G_p)^2 \right)$ \Comment{$O(d)$}
		\EndFor
		\State \textbf{return} $(G_\theta, G_p)$
		\EndFunction
	\end{algorithmic}
\end{algorithm}
\section{Convergence of NRPT chain} \label{app:NRPT_correctness}

\begin{figure}[t]
  \centering
  \begin{subfigure}{0.48\textwidth}
    \centering
    \includegraphics[width=0.95\textwidth]{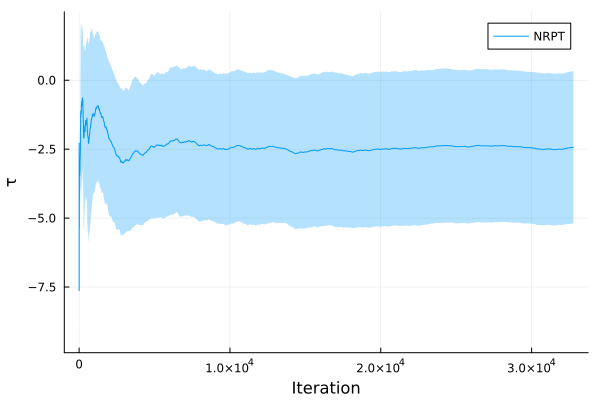}
  \end{subfigure}
  \begin{subfigure}{0.48\textwidth}
    \centering
    \includegraphics[width=0.95\textwidth]{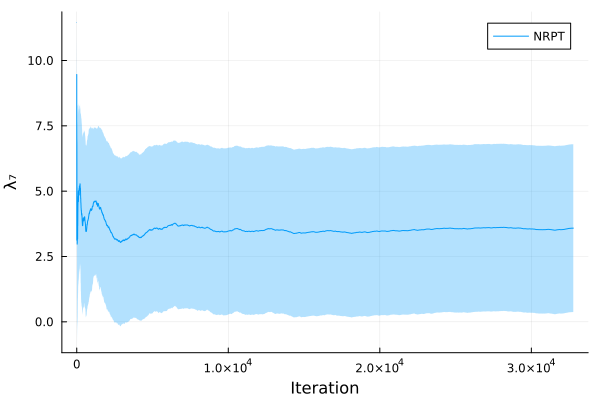}
  \end{subfigure}
  \begin{subfigure}{0.48\textwidth}
    \centering
    \includegraphics[width=0.95\textwidth]{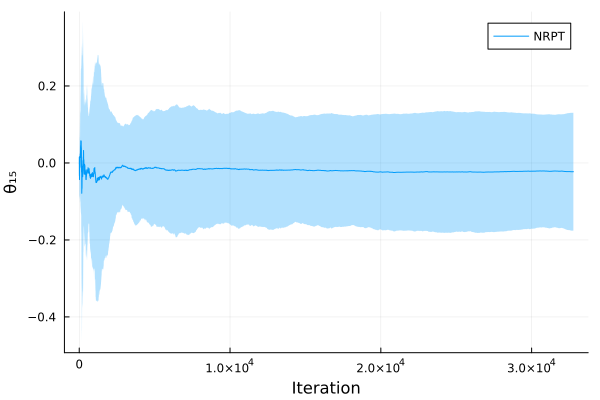}
  \end{subfigure}
  \begin{subfigure}{0.48\textwidth}
    \centering
    \includegraphics[width=0.95\textwidth]{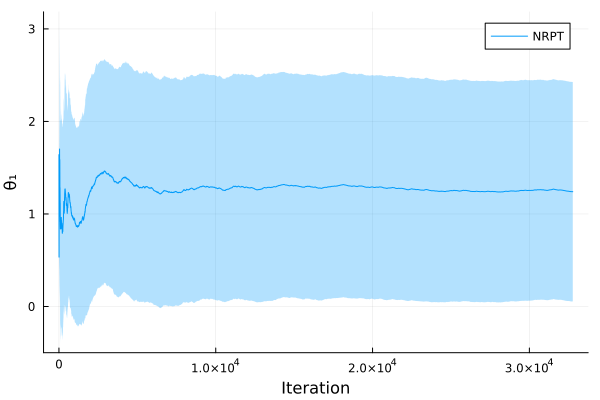}
  \end{subfigure}
  \caption{Running quartile plots of different parameters obtained from the 
  NRPT chain for the horseshoe logistic regression model on prostate cancer data 
  in \cref{sec:realdata_experiment}. The shaded regions are running interquartile ranges
  and the solid lines are running medians.}
  \label{fig:NRPT_trace}
\end{figure}

\begin{figure}[t]
  \centering
  \includegraphics[width=0.7\textwidth]{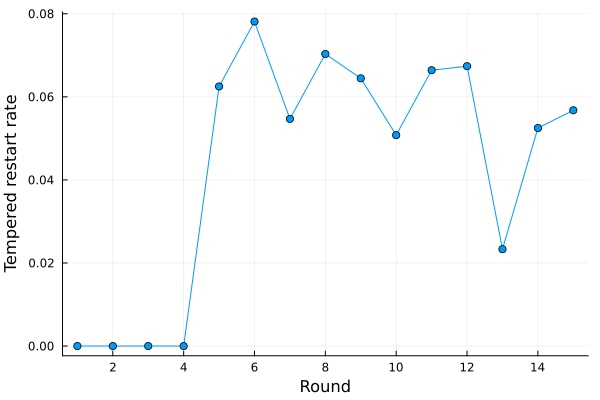}
  \caption{Restart rate over adaptation rounds of the NRPT chain for 
  the horseshoe logistic regression model 
  on prostate cancer data in \cref{sec:realdata_experiment}.}
  \label{fig:NRPT_restart}
\end{figure}

In this section, we show trace plots of the NRPT chain for the subsampled prostate
cancer data experiment in \cref{sec:realdata_experiment}
for different model parameters and the chain's tempered restart rate \citep{syed2022non} through each round of 
adaptation. In \cref{fig:NRPT_trace}, we see that the running quartiles of the chain
have stabilized for all parameters plotted. In \cref{fig:NRPT_restart}, we also
see that the tempered restart rate (the rate at which a sample from the reference chain 
percolates to the target chain) has stabilized around 0.06 starting from the fifth 
round of adaptation. These plots all indicate that the NRPT chain has reached 
stationarity and is suitable to be used as a reference posterior.

\appendix

\end{document}